\documentclass[a4paper,fleqn,usenatbib]{mnras}

\usepackage{mathptmx}

\usepackage[T1]{fontenc}
\usepackage{ae,aecompl}

\usepackage{graphicx}	
\usepackage{amsmath}	
\usepackage{amssymb}	
\usepackage{wasysym}           
\usepackage{epstopdf}
\usepackage{mathrsfs}
\usepackage{natbib}
\usepackage{physymb}
\usepackage{color}

\newcommand{\ed}[1]{{}{\color{black}#1}}

\title[Post-periapsis pancakes]{Post-periapsis pancakes: sustenance for self-gravity in tidal disruption events}

\author[Coughlin et al.]{
Eric R. Coughlin,$^{1,2}$\thanks{email: eric.coughlin@colorado.edu}
Chris Nixon,$^{1,3}$\thanks{Einstein fellow}
Mitchell C. Begelman,$^{1,2}$
Philip J. Armitage,$^{1,2}$
\newauthor
and Daniel J. Price$^{4}$
\\
$^{1}$JILA, University of Colorado and National Institute of Standards and Technology, 440 UCB, Boulder, CO 80309-0440, USA \\
$^{2}$Department of Astrophysical and Planetary Sciences, University of Colorado, 391 UCB, Boulder, CO 80309-0391, USA\\
$^{3}$Department of Physics \& Astronomy, University of Leicester, Leicester LEI 7RH UK \\
$^{4}$Monash Centre for Astrophysics, School of Physics \& Astronomy, Monash University, Clayton, Vic 3800, Australia
}

\date{Accepted XXX. Received YYY; in original form ZZZ}

\pubyear{2015}

\begin{document}
\label{firstpage}
\pagerange{\pageref{firstpage}--\pageref{lastpage}}
\maketitle

\begin{abstract}
A tidal disruption event, which occurs when a star is destroyed by the gravitational field of a supermassive black hole, produces a stream of debris, the evolution of which ultimately determines the observational properties of the event. Here we show that a post-periapsis caustic -- a location where the locus of gas parcels comprising the stream would collapse into a two-dimensional surface if they evolved solely in the gravitational field of the hole -- occurs when the pericenter distance of the star is on the order of the tidal radius of the hole. It is demonstrated that this ``pancake'' induces significant density perturbations in the debris stream, and, for stiffer equations of state (adiabatic index $\gamma \gtrsim 5/3$), these fluctuations are sufficient to gravitationally destabilize the stream, resulting in its fragmentation into bound clumps. The results of our findings are discussed in the context of the observational properties of tidal disruption events.
\end{abstract}




\section{Introduction}

A supermassive black hole of mass $M_h$ can tidally destroy a star of mass $M_*$ and radius $R_*$ if the star comes within the tidal radius $r_t \simeq R_*(M_h/M_*)^{1/3}$ of the hole. In this scenario, called a tidal disruption event (TDE), the star is shredded into a stream of debris. The properties of the debris and its ultimate fate have been studied for decades, both analytically and numerically, and the observational predictions generated from these studies have been tested.

Early analyses of TDEs showed that, due to the differential gravitational potential of the black hole, half of the disrupted debris that was closer to the hole at the time of disruption is bound to the black hole, while the other half is unbound \citep{lac82, ree88}. The half that is bound will eventually return to the black hole, circularize, and form an accretion disk. The properties and observational signatures of this accretion disk have been investigated by many authors (e.g., \citealt{can90, loe97, str09, str11, lod11, gui14b, she14, cou14}). The power radiated during the accretion process is enough to generate a highly luminous event, and some of these events have already been observed \citep{bad96, kom99, hal04, lev11, cen12, bog14, kom15}.

\citet{phi89} showed analytically that the rate at which the debris returns to the black hole decreases with time as $\dot{M}_{fb} \propto t^{-5/3}$. This feature, coupled with the longevity of the signature, is the observational ``smoking gun'' of a TDE. Many of the recently-observed TDE candidates exhibit a lightcurve that decreases in a manner commensurate with this power-law rate \citep{blo11, zau11, cen12, gez12, bog14, bro15}. 

To investigate the complex hydrodynamical interactions that take place during TDEs, many authors have resorted to numerical simulations. Early smoothed-particle hydrodynamics (SPH) calculations supported the analytic estimates of \citet{ree88} and \citet{phi89}, showing that the distribution of specific energies calculated not long after the time of disruption generates a fallback rate that scales as $\dot{M}_{fb} \propto t^{-5/3}$ \citep{eva89}. More recently, \citet{lod09} elucidated the effects of the structure of the progenitor star on the disruption process, demonstrating that the early stages of the fallback depend on the properties of the star. \citet{gui13} investigated how the impact parameter $\beta \equiv r_t/r_p$, $r_p$ being the pericenter distance of the stellar progenitor, alters the nature of the event, and found that shallower impact parameters often result in the survival of a bound stellar core. Finally, \citet{hay13}, \citet{bon15}, \citet{hay15}, and \citet{shi15} have looked into the effects of general relativity on the stream, showing how apsidal and Lense-Thirring precession can alter the formation of the disk that forms when the tidally-disrupted debris returns to pericenter. 

\citet{cou15} demonstrated that, when a solar-like star with a $\gamma = 5/3$ adiabatic equation of state is disrupted by a $10^6M_{\astrosun}$ hole, self-gravity can be important for determining the stream properties during its late evolution {}{(see also \citealt{koc94} and \citealt{gui14b} for a discussion of self-gravity)}. In particular, they showed that the tidal influence of the black hole becomes sub-dominant to the self-gravity of the debris, which results in the late-time fragmentation of the stream into gravitationally-bound clumps. These clumps then return to the original pericenter at discrete times, causing the fallback rate of the material to fluctuate about the $t^{-5/3}$ average.

An important question arising from the results of \citet{cou15} is: when is the self-gravitational nature of the stream revived post-disruption? As the tidal shear and the self-gravity of the star equal one another at the tidal radius, one might suspect that the self-gravity of the debris is most influential at late times. Indeed, it is during this late evolution that \citet{cou15} found that the stream gravitationally fragments. As we will show here, however, the self-gravity of the debris can affect the stream evolution soon after the star passes through periapsis (on the order of hours for the tidal disruption of a solar-like star by a $10^6M_{\astrosun}$ hole). We find that the star experiences compressive forces in the orbital plane, which lead to the formation of a post-disruption pancake, similar to the one found by \citet{car82} but oriented orthogonal to the orbital plane of the progenitor. This in-plane recompression then augments the importance of self-gravity, resulting in perturbations on top of the stream that can induce early recollapse. (We note that we will be considering TDEs in which the star is completely destroyed, and hence these results should not be confused with those of \citealt{gui13} who, in certain cases, found surviving stellar cores for impact parameters less than one.)

In Section 2 we present an analytical analysis of the stream under the impulse approximation, which assumes that the star maintains hydrostatic balance until it reaches the tidal radius. We demonstrate that, even when the pericenter distance and the tidal radius are approximately coincident, a caustic -- a location where the orbits of the gas parcels comprising the stream collapse to a two-dimensional surface -- occurs shortly after the star is disrupted. Section 3 presents numerical simulations that demonstrate the effects of this caustic, and specifically shows how it can modify the density structure of the stream for times long after disruption. We present a discussion of the results of our simulations in Section 4 and  consider the astrophysical implications of our findings in Section 5. We conclude and summarize in Section 6.

\section{The impulse approximation}
Many authors (e.g., \citealt{car83,ree88,lod09,sto13,cou14}) have considered the disruption process from a simplified, analytic standpoint. While an analytic approach almost certainly misses many of the intricacies associated with the realistic problem, it has the advantage of being able to characterize the bulk processes that take place during the interaction. Furthermore, it is able to elucidate the manner in which those processes depend on the properties of the progenitor star and the black hole, which provide useful observational diagnostics. Here we discuss the impulse approximation, which assumes that the star is able to maintain hydrostatic balance until it reaches the tidal radius and it is thereafter disrupted, i.e., the pressure and self-gravity of the material are negligible after the star has passed through the tidal radius.

\citet{car82, car83} considered the case where the pericenter distance of the star, $r_p$, is well inside the tidal radius of the hole (their affine star model; {}{see also \citealt{sto13} for an alternative approach to analytically modeling this scenario}). For these high-$\beta$ encounters, where $\beta \equiv r_t/r_p$ is the impact parameter, the impulse approximation can be applied early on in the tidal disruption process. Because of the component of the tidal force that acts orthogonally to the orbital plane of the star, the gas parcels comprising the top and bottom of the stellar envelope undergo effective freefall, forming an infinitely thin plane, or caustic, at the pericenter radius (the location of the caustic is actually slightly after the pericenter, only equaling the pericenter distance for $\beta \rightarrow \infty$; \citealt{bic83}). This ``pancaking'' effect was then thought to be capable of igniting thermonuclear fusion via the triple-$\alpha$ process, resulting in the detonation of the star. However, studies showed that the shocks near pericenter resulted in lower densities and pressures in the stellar core than those predicted by \citet{car82}, meaning that the triple-$\alpha$ process is unlikely to be initiated in these encounters (though some fusion via the CNO cycle may occur; \citealt{bic83}).

On the other hand, when $\beta \simeq 1$, the star can retain its unperturbed structure for much longer. In this case, one can approximate the star as being spherical, with every gas parcel moving with the center of mass, until the pericenter is reached. Here we will focus on this case, not only because it has not been treated as thoroughly as the $\beta \gg 1$ scenario, but also because it has interesting consequences for the disrupted material soon after the pericenter distance is reached. Later in this paper, we will relax the assumptions made by this model with three dimensional hydrodynamic simulations.

\subsection{Equations}
Once the star passes through the tidal radius, the {}{pressure and self-gravity of the} gas parcels are assumed {}{negligible, implying} that they follow Keplerian orbits in the potential of the black hole. The equations of motion that describe these orbits are given by

\begin{equation}
r^2\sin^2\theta\,\dot\phi = \ell \label{rkep1},
\end{equation}
\begin{equation}
r^4\dot\theta^2+\frac{\ell^{\,2}}{\sin^2\theta} = k^2 \label{rkep2},
\end{equation}
\begin{equation}
\frac{1}{2}\bigg{(}\dot{r}^2+\frac{k^2}{r^2}\bigg{)}-\frac{GM}{r} = \epsilon \label{rkep3},
\end{equation}
where dots denote differentiation with respect to time, $r(t)$, $\theta(t)$, and $\phi(t)$ are the respective radial, polar, and azimuthal coordinates of the gas parcel under consideration, and $M$ is the mass of the black hole. Here $\ell$, $k$, and $\epsilon$ are constants of integration, the first two being projections of the specific angular momentum, while the last is the specific energy.

\begin{figure}
   \centering
   \includegraphics[width=3.4in]{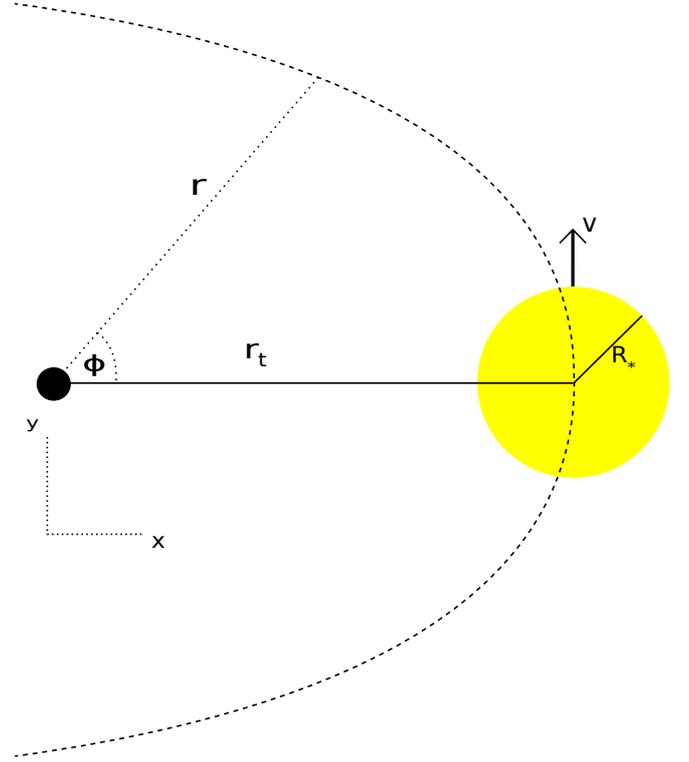} 
   \caption{The initial configuration of the star under the impulse approximation when $\beta \simeq 1$ (this figure is not drawn to scale). The dashed curve traces out the orbit of the center of mass, which is assumed to be parabolic. {}{The Cartesian coordinates are indicated by the diagram immediately below the black hole (which is indicated by the black circle), $z$ being out of the plane in a right-handed sense. The spherical-polar coordinates are labeled $r$ and $\phi$ on the diagram, and $\theta$ is measured out of the plane of the orbit from the z-axis (for the above figure that focuses on the $x$-$y$ plane, $\theta = \pi/2$).}}
   \label{fig:tde_setup}
\end{figure}

Setting the impact parameter to $\beta \equiv r_t/r_p = 1$, the point at which equations \eqref{rkep1} -- \eqref{rkep3} become valid occurs when the star reaches pericenter. We will let the orbit of the stellar progenitor be confined to the $xy$-plane, with the periapsis on the positive-$x$ axis and the location of the black hole at the origin. The center of mass of the star will also trace out a parabolic orbit. The impulse approximation then means that the star retains its unperturbed (assumed-spherical) structure until it reaches pericenter, so that the initial conditions we will use for equations \eqref{rkep1} -- \eqref{rkep3} will be those depicted by Figure \ref{fig:tde_setup}. Note that the entire star initially shares the velocity of the center of mass, which is along the positive-$y$ axis at pericenter.

With the setup given by Figure \ref{fig:tde_setup} in mind, we will define the initial position of a given fluid element that comprises the star by the coordinates $(r_{i},\,\theta_{i},\,\phi_i)$. Since the entire star moves with the center of mass, the velocity of every fluid element is given by $\dot{z}_i = \dot{x}_i = 0$, $\dot{y}_i = \sqrt{2GM_h/r_t}$. Transforming these conditions into spherical coordinates via the transformations $z = r\cos\theta$, $y = r\sin\theta\sin\phi$, $x = r\sin\theta\cos\phi$, we find 

\begin{equation}
\dot{r}_i = \sqrt{\frac{2GM_h}{r_t}}\sin\theta_i\sin\phi_i \label{rdoti},
\end{equation}
\begin{equation}
\dot\theta_i = \frac{1}{r_i}\sqrt{\frac{2GM_h}{r_t}}\cos\theta_i\sin\phi_i,
\end{equation}
\begin{equation}
\dot\phi_i = \frac{1}{r_i}\sqrt{\frac{2GM_h}{r_t}}\frac{\cos\phi_i}{\sin\theta_i},
\end{equation}
and using these expressions in equations \eqref{rkep1} -- \eqref{rkep3} gives

\begin{equation}
\ell = r_i\sqrt{\frac{2GM_h}{r_t}}\sin\theta_i\cos\phi_i \label{elleq},
\end{equation}
\begin{equation}
k = r_i\sqrt{\frac{2GM_h}{r_t}}\sqrt{\cos^2\phi_i+\cos^2\theta_i\sin^2\phi_i} \label{keq},
\end{equation}
\begin{equation}
\epsilon = \frac{GM_h}{r_t}\left(1-\frac{r_t}{r_i}\right) \label{epseq}.
\end{equation}
Equation \eqref{epseq} shows that gas parcels with initial positions inside the tidal radius are bound ($\epsilon < 0$), while those outside are unbound ($\epsilon > 0$), which is what we expect. 

In addition to its position, we will also be interested in the density of the stream. As was demonstrated in \citet{cou15}, the density structure can be determined by considering the star at the time of disruption and assuming that the specific energies of the gas parcels are frozen in thereafter. Making the additional assumption that the stream is a circular cylinder of cross-sectional radius $H$, then we can show that the azimuthally-averaged density along the stream varies as \citep{cou15}

\begin{equation}
\rho = \frac{M_*\xi_1}{2\pi{}H^2\sqrt{(r')^2+r^2(\phi')^2}}\frac{\int_{\mu\xi_1}^{\xi_1}\Theta(\xi)^{n}\xi{}d\xi}{\int_0^{\xi_1}\Theta(\xi)^{n}\xi^2d\xi} \label{rhostream}, 
\end{equation}
where $M_*$ is the mass of the disrupted star, $n = 1/(\gamma-1)$ is the polytropic index of the gas, $\Theta(\xi)$ is the solution to the Lane-Emden equation and $\xi_1$ is the first root of $\Theta(\xi)$ \citep{han04}. Here $\mu$ is the dimensionless position of a gas parcel from the center of the star at the time of disruption, i.e., $\mu = R_p/R_*$, where $R_p$ is the radial position of the gas parcel. Primes on the functions $r$ and $\phi$ denote differentiation with respect to $\mu$. We will return to the question of what determines $H$ in Section 3.2.

\subsection{Solutions}
With equations \eqref{elleq} -- \eqref{epseq} and the initial positions of the gas parcels, we can numerically integrate equations \eqref{rkep1} -- \eqref{rkep3} to determine the temporal evolution of the debris stream. 

\begin{figure}
   \centering
   \includegraphics[width=3.4in]{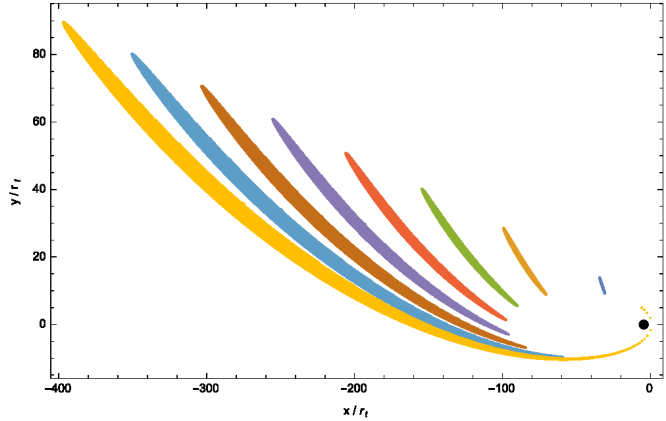} 
   \caption{The streams of debris formed from the tidal disruption of a solar-type star by a $M_h = 10^6M_{\astrosun}$ hole situated at the origin. Each color represents a different time, the earliest (blue points closest to the origin) being at $t = 100\,r_t^{3/2}/\sqrt{GM_h} \simeq 1.84$ days from disruption, the latest (yellow points) at $t = 2200\,r_t^{3/2}/\sqrt{GM_h} \simeq 40.6$ days from disruption. The time in between neighboring streams is $300\,r_t^{3/2}/\sqrt{GM_h} \simeq 5.53$ days. The black hole (not drawn to scale) is indicated by the black circle near the origin. {}{While the radial positions of the gas parcels match well those from numerical analyses (see Figure 1 of \citealt{cou15}), the width obtained from equations \eqref{rkep1} -- \eqref{rkep3} is significantly overestimated (the numerical solutions, had we shown them, would have amounted to lines plotted overtop of the streams in Figure 2). This finding suggests that self-gravity is important for keeping the stream confined in the transverse direction.}}
   \label{fig:tidal_streams}
\end{figure}

Figure \ref{fig:tidal_streams} shows the solution to equations \eqref{rkep1} -- \eqref{rkep3} with the relevant initial conditions for a TDE between a solar-type star and a $10^6M_{\astrosun}$ hole. The first time (closest set of blue points) is 1.84 days after disruption, while the longest, yellow set of points is 40.6 days after disruption, and coincides roughly with the time at which the most bound material has returned to pericenter. Intermediate streams are shown at intervals of 5.53 days. We find overall good qualitative and quantitative agreement between the radial positions of these solutions and the solution to the full problem -- making no assumption about the negligible nature of pressure and self-gravity -- obtained using numerical simulations (see, e.g., the red curves in Figure 1 of \citealt{cou15}).

\begin{figure*}
   \centering
   \includegraphics[width=.495\textwidth, height=.495\textwidth]{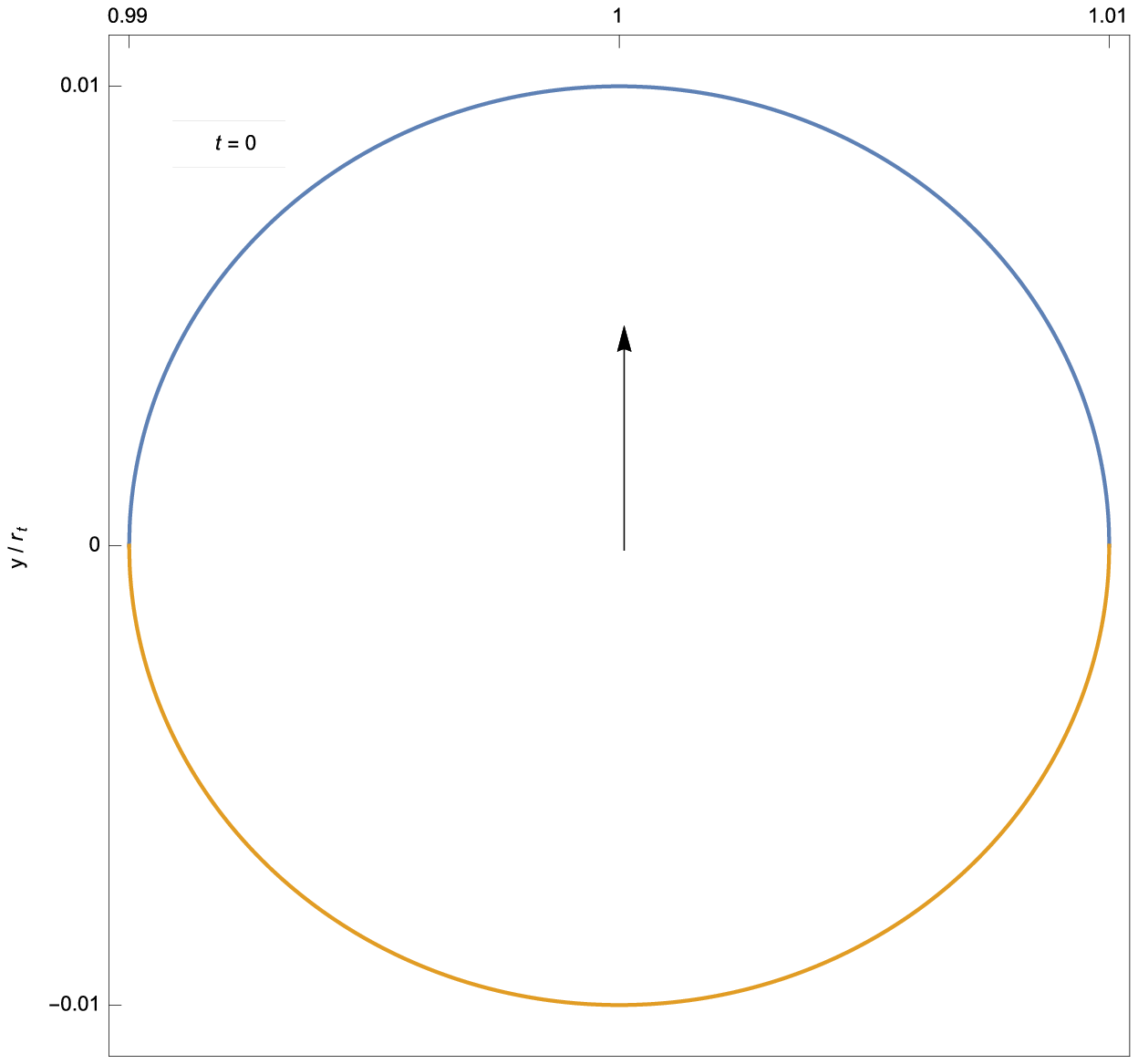}  
     \includegraphics[width=.48\textwidth, height=.475\textwidth]{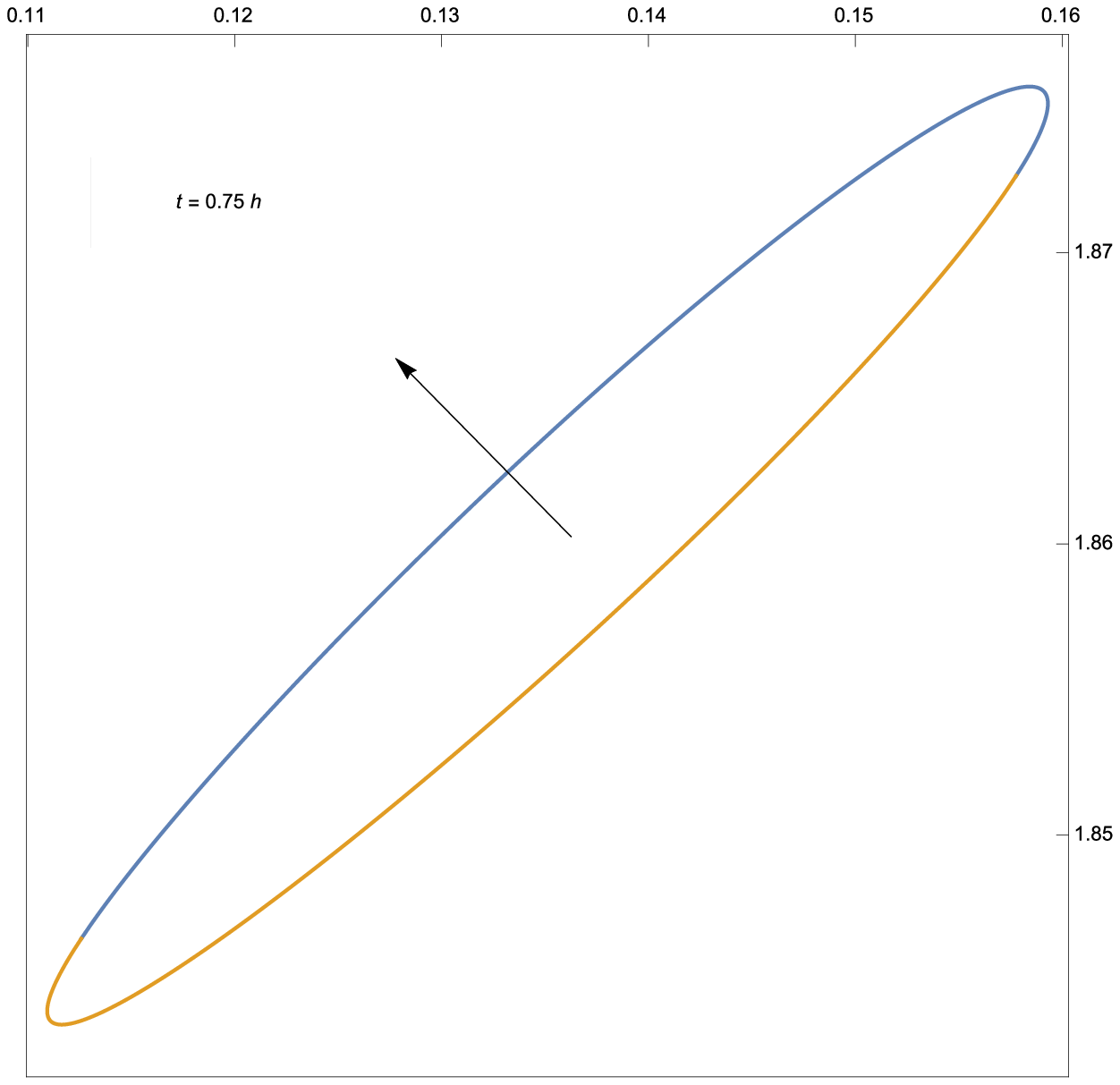}
     \hspace{3pt}
    \includegraphics[width=.495\textwidth, height=.495\textwidth]{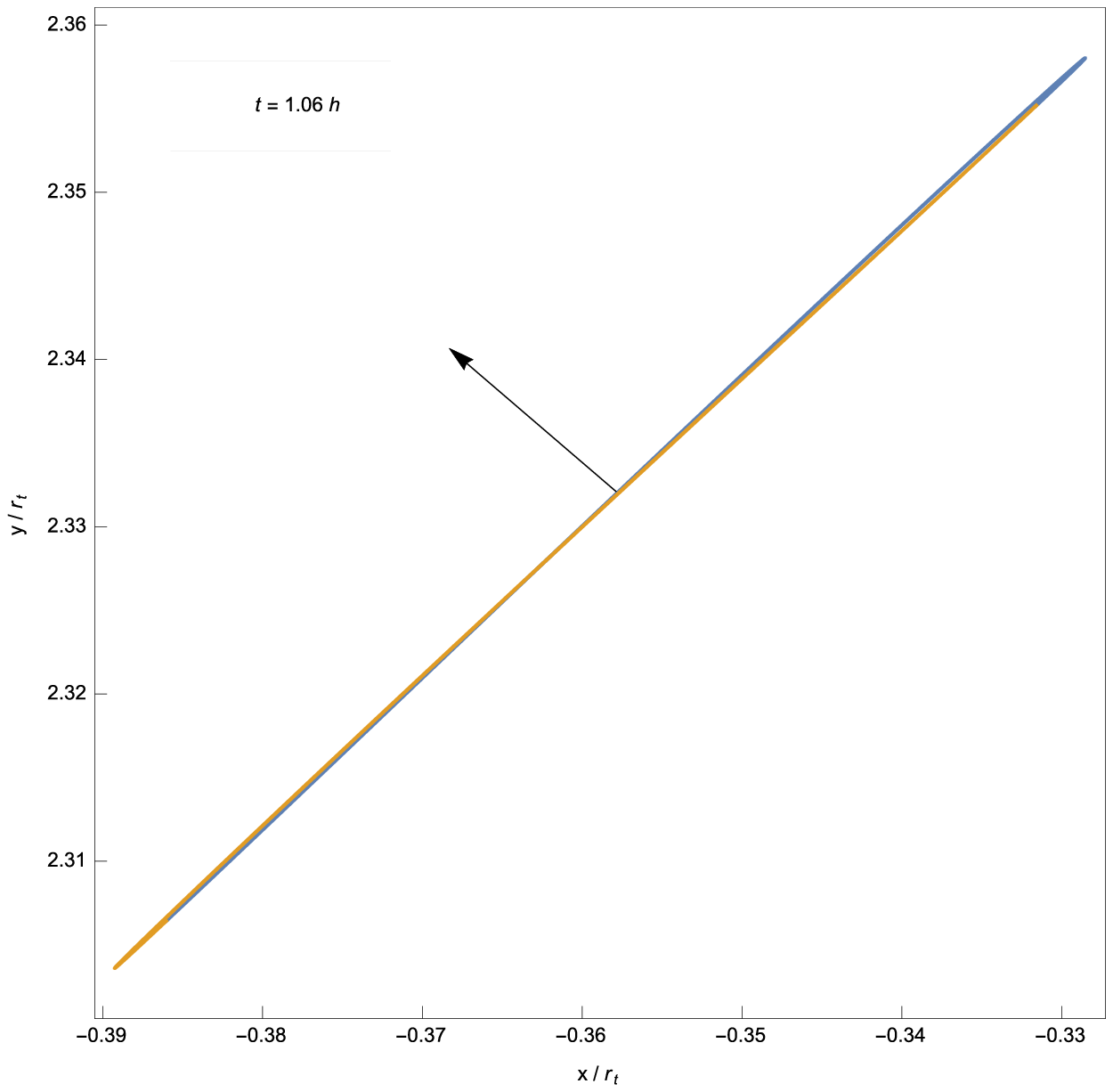}
    \hspace{3.pt}  
        \includegraphics[width=.485\textwidth, height=.485\textwidth]{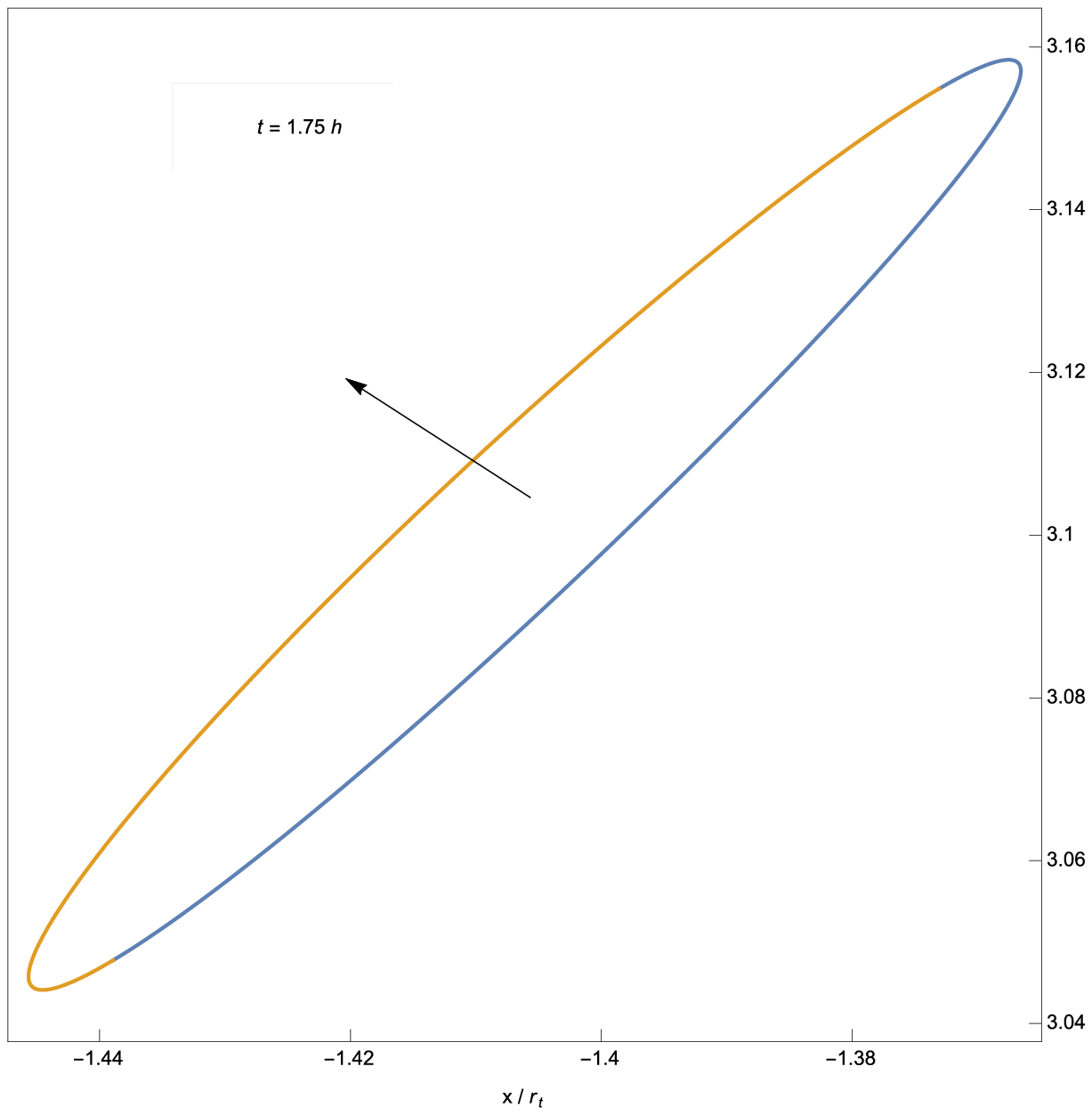} 
   \caption{Four snapshots of the in-plane evolution of the gas parcels comprising the edge of the star at the time of disruption; for these figures we chose a $10^6M_{\astrosun}$ hole and a solar-like star. The particles comprising the front of the star at the time of disruption have been colored blue, while the back has been colored orange. {}{The arrows indicate the direction of motion of the center of mass.} The bottom, left-hand panel shows that, at a time of roughly an hour after disruption, the front and back of the stream merge and thereafter trade places. The impulse approximation thus leads to a caustic -- where the debris streams form a two-dimensional surface -- which occurs roughly an hour after disruption.}
   \label{fig:pancakes}
\end{figure*}

However, we find disagreement between the width of the stream obtained from equations \eqref{rkep1} -- \eqref{rkep3} and that from the simulations, the former being significantly wider than the latter. This discrepancy is due to the fact that self-gravity plays a crucial role in determining the width of the stream \citep{cou15}. In other words, the $H$ that appears in equation \eqref{rhostream} is not simply determined by the free expansion of the parcels in the gravitational potential of the hole (see equation \eqref{Hscale} below, which shows how $H$ depends on the density of the stream in the limit that hydrostatic balance is upheld in the transverse direction). 

The approximate point at which the self-gravity of the stream becomes important can, however, be gleaned from the solutions to equations \eqref{rkep1} -- \eqref{rkep3}. Figure \ref{fig:pancakes} shows the evolution of the in-plane edge of the stream at four different times for the disruption of a solar-type star by a $10^6M_{\astrosun}$ hole. The front of the stream (the fluid parcels comprising the leading edge of the polytrope at the time of disruption) has been colored blue, the back has been colored orange{}{, and the arrow indicates the instantaneous direction of motion of the center of mass}. This figure demonstrates that, roughly an hour after disruption, the leading and trailing edges of the stream form a caustic -- a point where the two-dimensional, in-plane surface of the stream collapses to a one-dimensional line -- and thereafter trade places, the front becoming the back and the back becoming the front.

The tidal stream thus exhibits a ``perpendicular pancake'' shortly after disruption, the perpendicular aspect referring to the fact that the orientation of the pancake is orthogonal to the orbital plane of the debris. This pancake is analogous to but distinct from the one found by \citet{car82}, who noted that the top and bottom of the star flatten to a point of infinite density at the tidal radius for high-$\beta$ encounters. Here, however, the compressive motions occur in the orbital plane. 

The existence of the pancake encountered here can ultimately be attributed to the initial conditions: from Figure \ref{fig:tde_setup}, it is apparent that the parcels along the line passing through the center of the star and perpendicular to the orbital plane all have their periapses at $\phi = 0$. Those constituting the leading edge of the star, however, have already passed through their periapses, while the periapses of the parcels comprising the back of the star have not yet been reached. From the conservation of angular momentum \eqref{rkep1}, the front of the star is therefore decelerating at the time of disruption while the back is accelerating, which causes the two to cross at a certain location. Specifically, if we differentiate equation \eqref{rkep1} with respect to time, set $\theta = \pi/2$ and use equation \eqref{rdoti}, we find

\begin{equation}
\ddot{\phi_i} = -\frac{4GM_h}{r_i^2r_t}\sin\phi_i\cos\phi_i \label{phiddot},
\end{equation}
which shows that gas parcels with $\phi_i > 0$ are decelerating in the $\phi$ direction, while those with $\phi_i < 0$ are accelerating. Investigating this equation further, we see that the differential acceleration across the star at the time of disruption is

\begin{equation}
\Delta{\ddot{\phi}} \simeq -\frac{4GM_h}{r_i^2r_t}\Delta{\phi},
\end{equation}
where $\Delta{\phi}$ is the angle subtended by the star. Geometrically $\Delta{\phi} \simeq 2R_*/r_t$, which yields, after setting $r_i \simeq r_t$, 

\begin{equation}
|\Delta{\ddot{\phi}}| \simeq \frac{32\pi{G}\rho_*}{3}\left(\frac{M_h}{M_*}\right)^{-1/3} \label{Deltaphidot},
\end{equation}
where $\rho_* = 3M_*/(4\pi{R_*}^3)$ is the average stellar density. This expression shows that the change in acceleration from the front of the star to the back depends primarily on the properties of the progenitor{}{, though the inverse dependence on black hole mass shows that the effect should be amplified for smaller-mass SMBHs}.

During a realistic $\beta \simeq 1$ tidal encounter, the star will not retain perfect spherical symmetry until reaching its pericenter. In particular, the outer, low-density material comprising the envelope will be more easily stripped, resulting in an elongated, ellipsoidal configuration. However, the higher-density core will be able to better maintain its structure. Therefore, while considering the entire star as spherical and moving with the center of mass at the time of disruption is likely too simplistic for the physical problem, those initial conditions are perhaps reasonable for the central regions. 

Furthermore, the non-zero pressure of the gas will prevent the development of a true caustic. On the contrary, the convergence of the Keplerian orbits will increase the pressure and density until it reaches an approximate equilibrium. However, the stretching of the stream in the radial direction will cause the density to decrease, which will likewise result  in a more drastic lowering of the pressure if the gas follows an adiabatic equation of state. The ability of the pressure to resist the caustic will thus decrease with time, making it possible for the perpendicular pancake to alter the nature of the debris stream.

{}{The precise time at which the caustic occurs as it has been presented here depends only on the gravitational field of the black hole. In reality, the self-gravity of the steam would serve to alter the precise nature of the pancake. However, we expect that self-gravity would only serve to enhance the focusing of the orbits and potentially generate the caustic at a slightly earlier time.}

In the next section we present simulations that address the complexity of the full problem. As we will see, the numerical solutions do exhibit interesting behavior near the time at which equations \eqref{rkep1} -- \eqref{rkep3} predict the existence of a caustic, and this behavior is imprinted on the stream for much later times.

\section{Numerical simulations}
To test whether or not the caustic discussed in the previous section affects realistic $\beta \simeq 1$ tidal encounters, we now employ numerical simulations that allow the star to evolve in the tidal field of the hole pre-periapsis and include the effects of pressure and self-gravity at all times. 

\subsection{Simulation setup and initial conditions}
We use the SPH code {\sc phantom} \citep{pri10, lod10} to simulate the tidal disruption of a solar-type star (one with a solar mass and a solar radius) by a $10^6M_{\astrosun}$ black hole. {\sc phantom} is a highly efficient code and is especially useful for astrophysical problems involving complex geometries and a large range of spatial and temporal scales. For other applications of this code, see, e.g., \citet{nix12a, nix12b, mar14a, mar14b, nea15}. 

In our simulations the star is initially assumed to be a polytrope with polytropic index $\gamma$ \citep{han04}. The correct, polytropic density profile is obtained by first placing $10^6$ particles in a close-packed sphere, then stretching that sphere to obtain a good approximation to the exact solution. 

We place the polytrope at a distance of $10\,r_t$ from the hole, with the center of mass on a parabolic orbit. The distance at periapsis is equal to the tidal radius ($\beta = 1$). Every gas parcel composing the star initially moves with the center of mass when the star is at $10\,r_t$, and the length of time taken to traverse the distance to the hole is sufficient to allow the polytrope to relax. {}{The adiabatic index of the gas is always equal to the initial, polytropic index of the star.}

Self-gravity is included at all stages of the TDE, and is employed via a k-D tree \citep{gaf11} alongside an opening angle criterion, the latter employing a direct summation method for the gravitational forces between neighboring particles \citep{pri07b}. The simulations presented here used an opening angle of $0.5$ (we have run simulations with smaller opening angles and found negligible differences; see \citealt{cou15}). Shock heating was not included for the runs presented here, though we have done tests in which it was included and found only negligible differences. We also do not account for non-adiabatic cooling; the gas therefore retains its polytropic equation of state throughout the TDE.  

We ran four different simulations, each identical to the next except in the adiabatic index used for the gas. Specifically, we chose $\gamma = 1.5,$ 5/3, 1.8, and 2, and thus our parameter space agrees with that chosen by \citet{lod09} except for $\gamma = 2$. While adiabatic indices greater than 5/3 are difficult to realize physically {}{in stellar progenitors (though they may be appropriate for planets; \citealt{fab05, li02})}, we included these cases to highlight the presence of the caustic and to compare to \citet{lod09}.

\subsection{Results}
\begin{figure*}
   \centering
   \includegraphics[width=.495\textwidth, height=.43\textwidth]{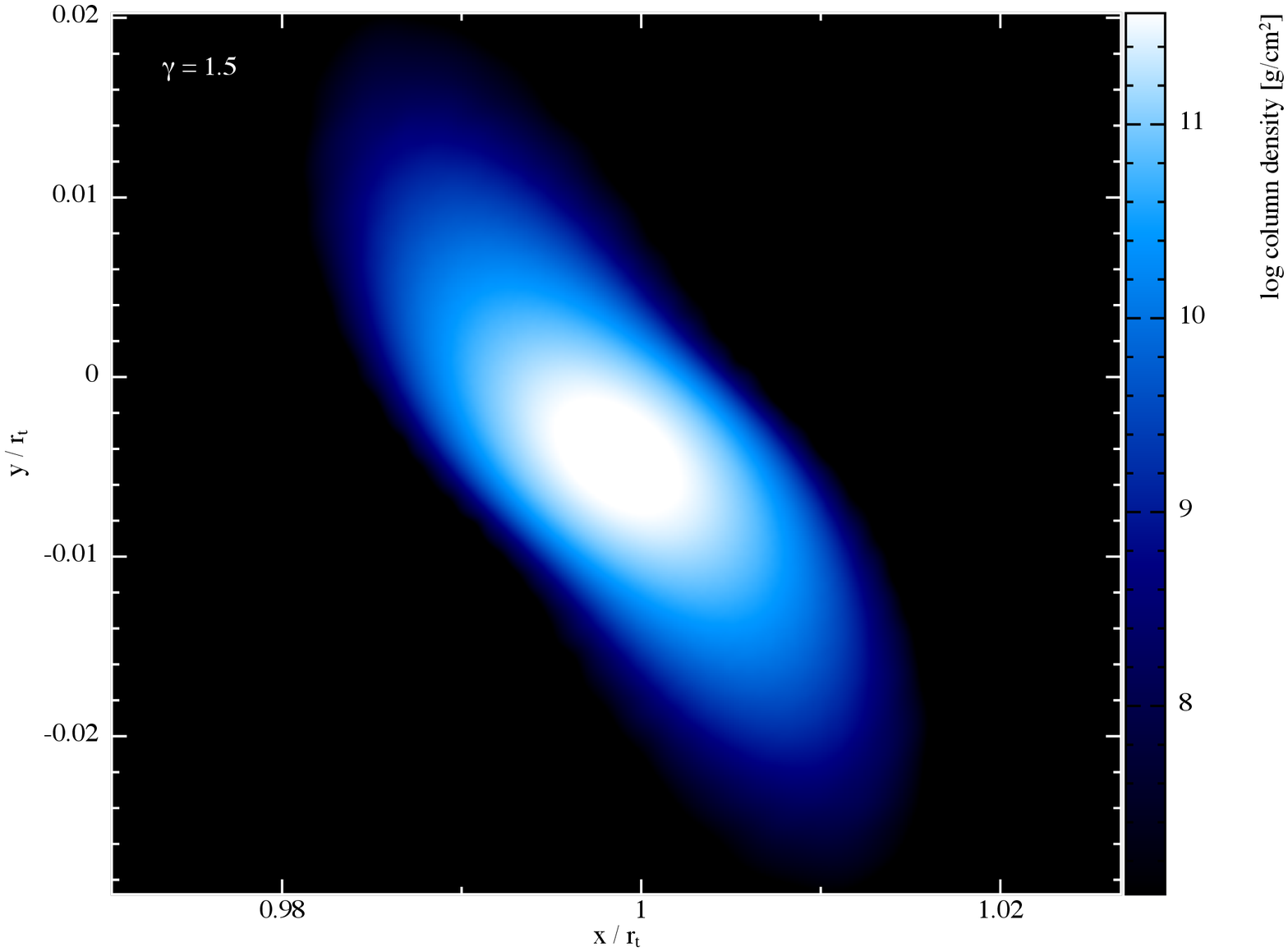} 
   \includegraphics[width=.495\textwidth, height=.43\textwidth]{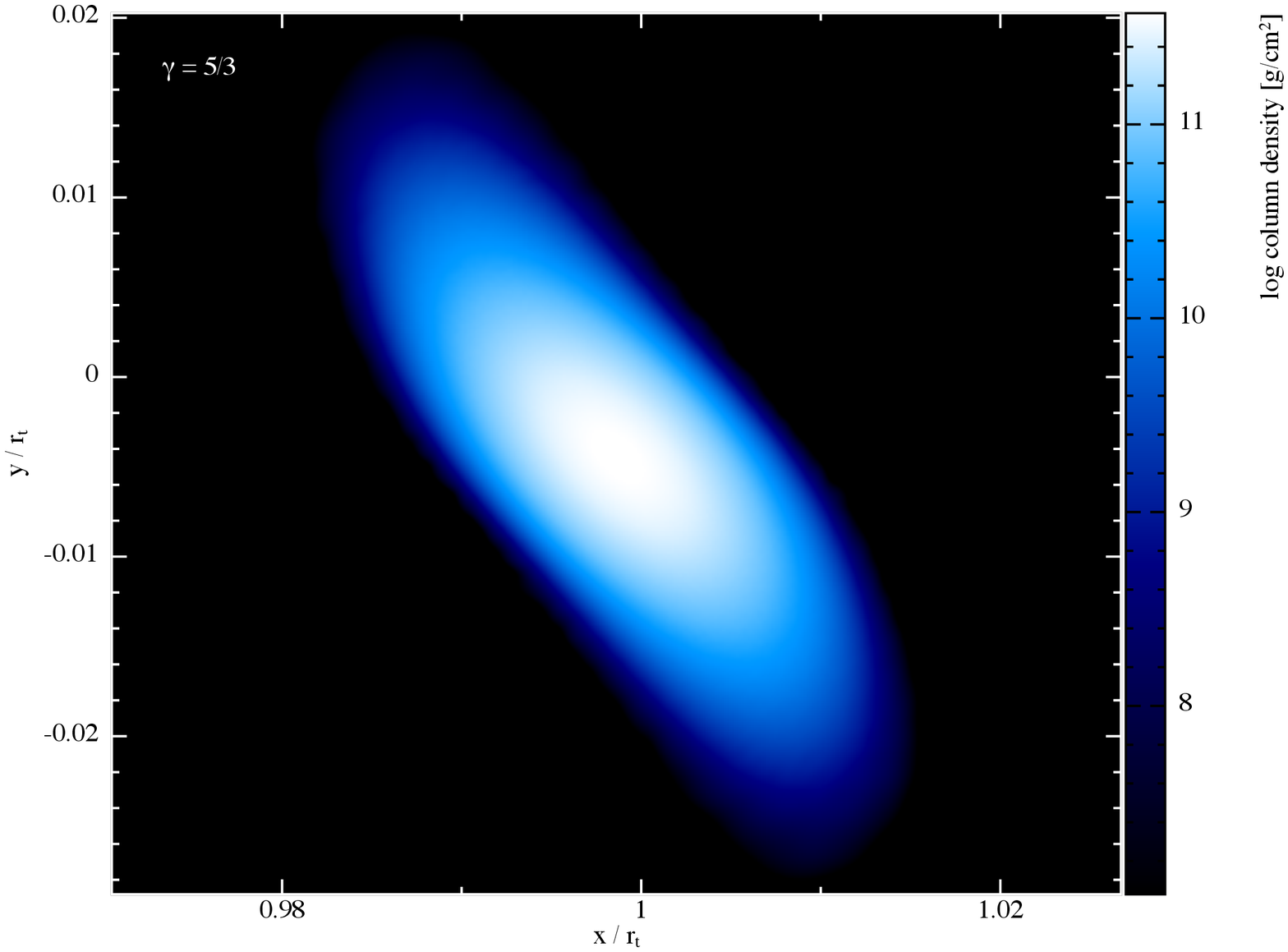}
   \includegraphics[width=.495\textwidth, height=.43\textwidth]{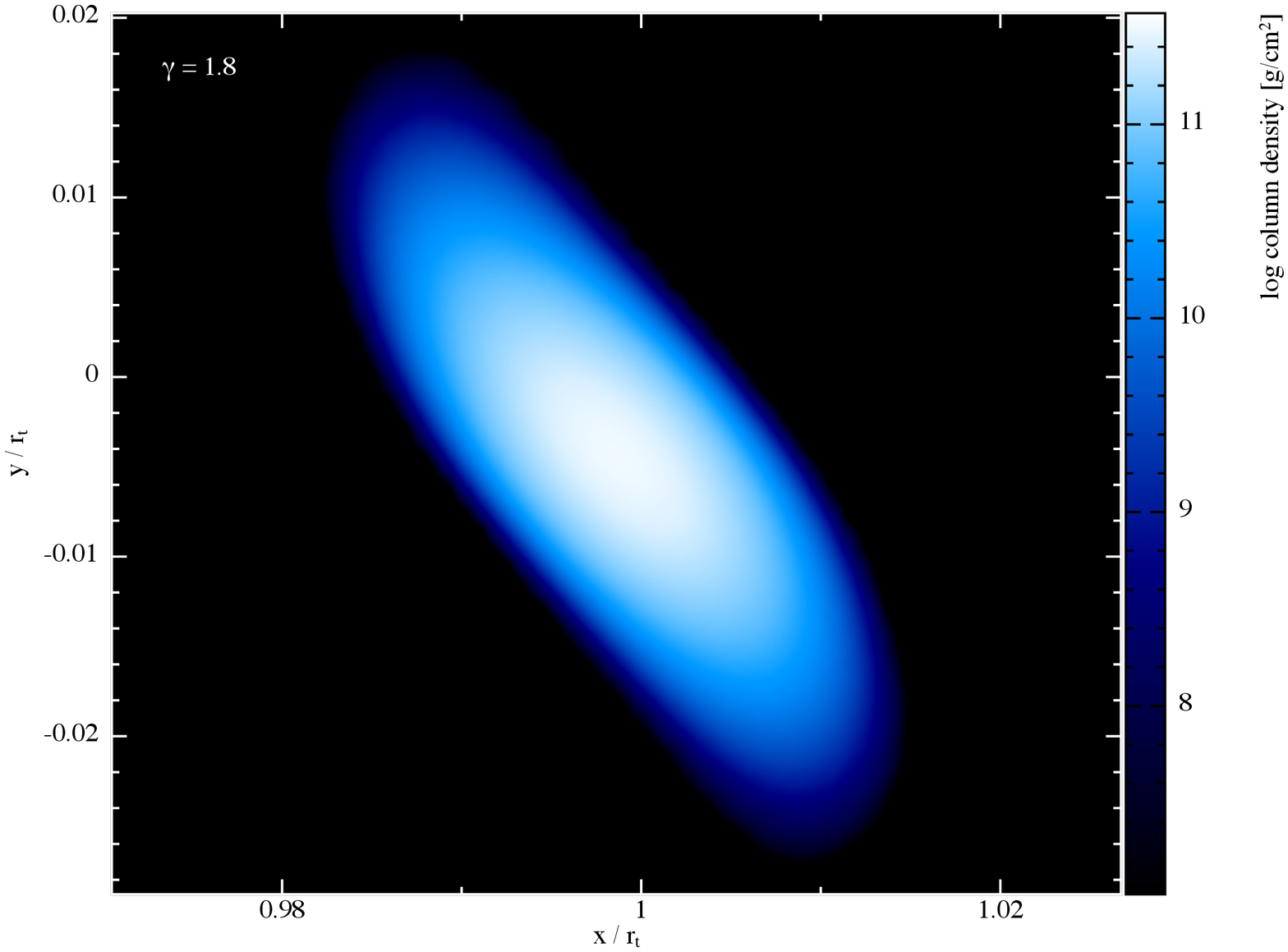} 
   \includegraphics[width=.495\textwidth, height=.43\textwidth]{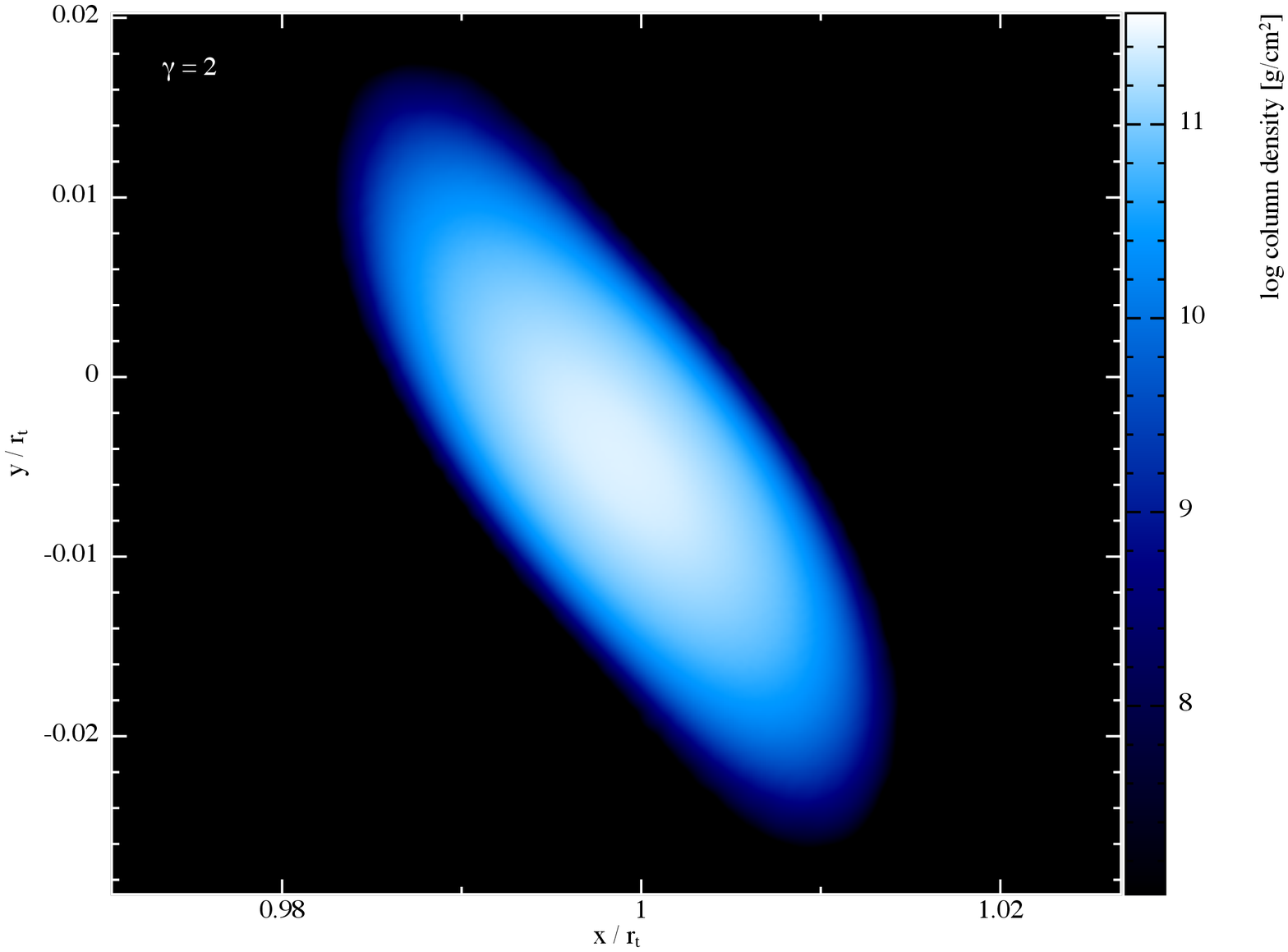} 
   \caption{The star at the time of disruption for an adiabatic index of $\gamma = 1.5$ (top, left), $\gamma = 5/3$ (top, right), $\gamma = 1.8$ (bottom, left) and $\gamma=2$ (bottom, right). The configuration has clearly been altered from its original, spherical shape, showing that the tidal force does not act exactly as an impulse as was assumed in Section 2. The central density is also higher for smaller $\gamma$, which is predicted from the original stellar profile.}
   \label{fig:todplots}
\end{figure*}

Figure \ref{fig:todplots} shows the star at the time of disruption, with each panel corresponding to a different adiabatic index. As was commented upon in Section 2, the fact that the tidal force does not act impulsively means that the polytrope is already distorted when it reaches its periapsis, and this distortion is apparent from the figure. We also see that the central density is higher for lower $\gamma$, which is a general feature of polytropes. 

\begin{figure*}
   \centering
   \includegraphics[width=.495\textwidth, height=.43\textwidth]{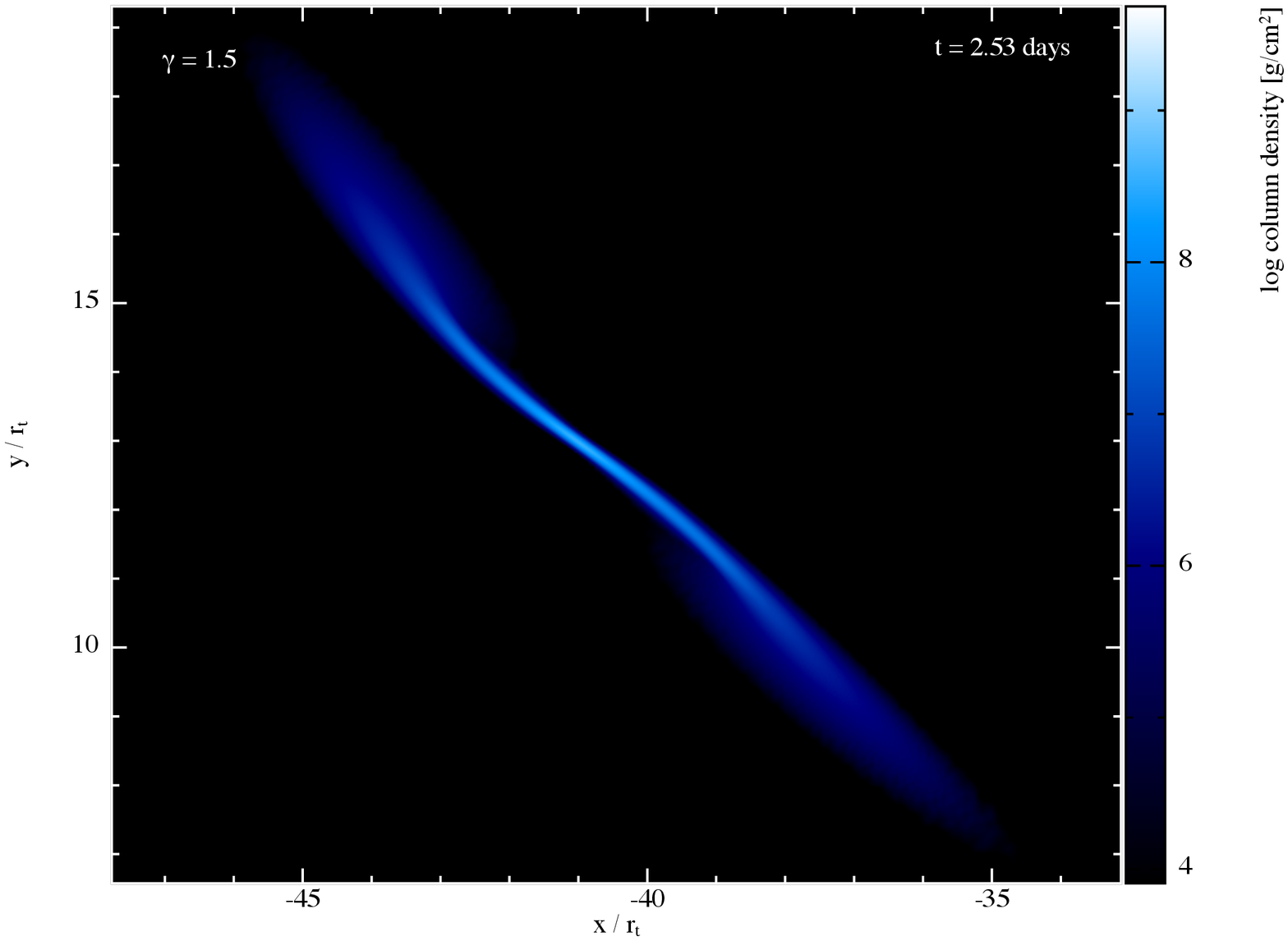} 
   \includegraphics[width=.495\textwidth, height=.43\textwidth]{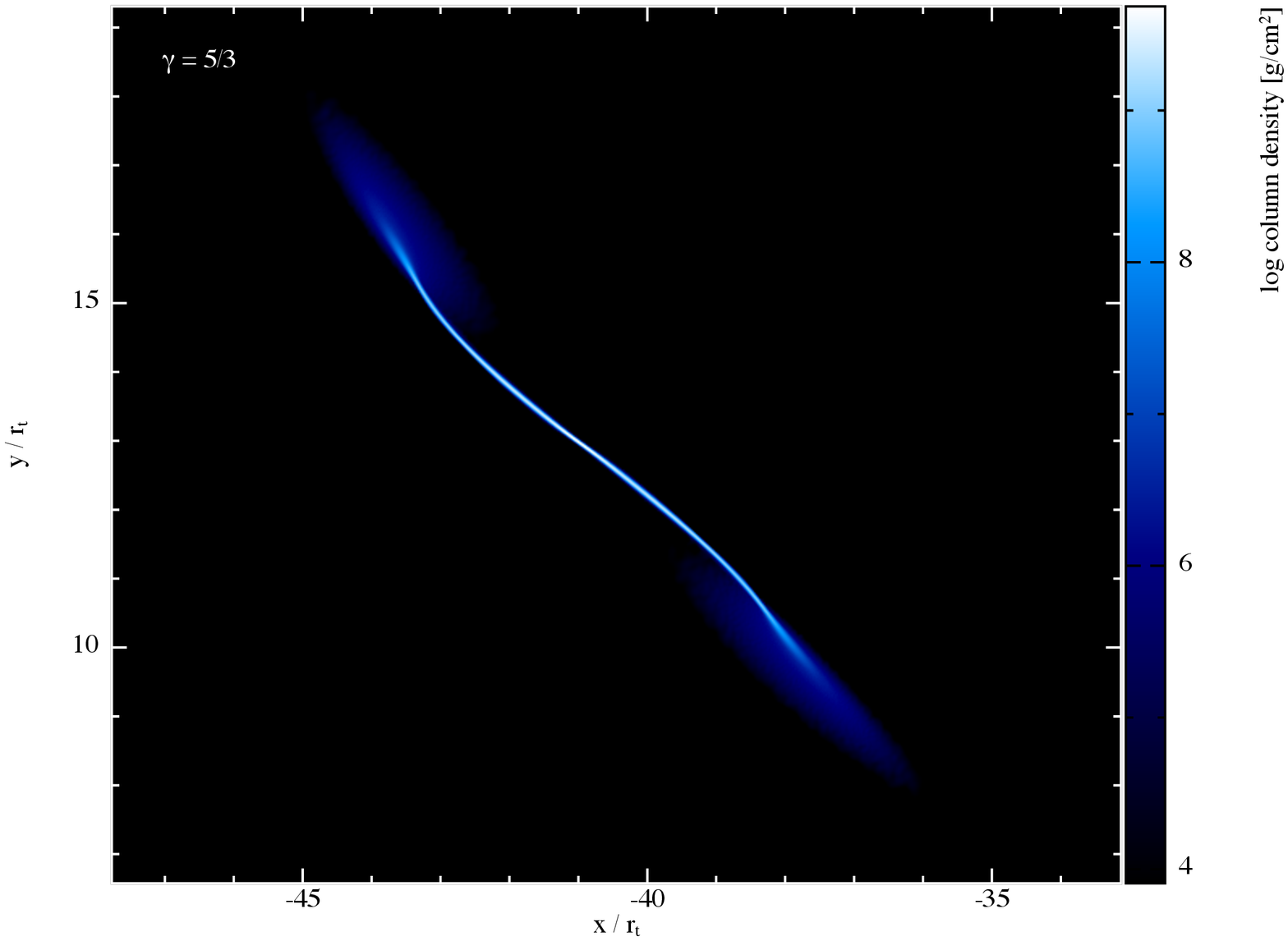}
   \includegraphics[width=.495\textwidth, height=.43\textwidth]{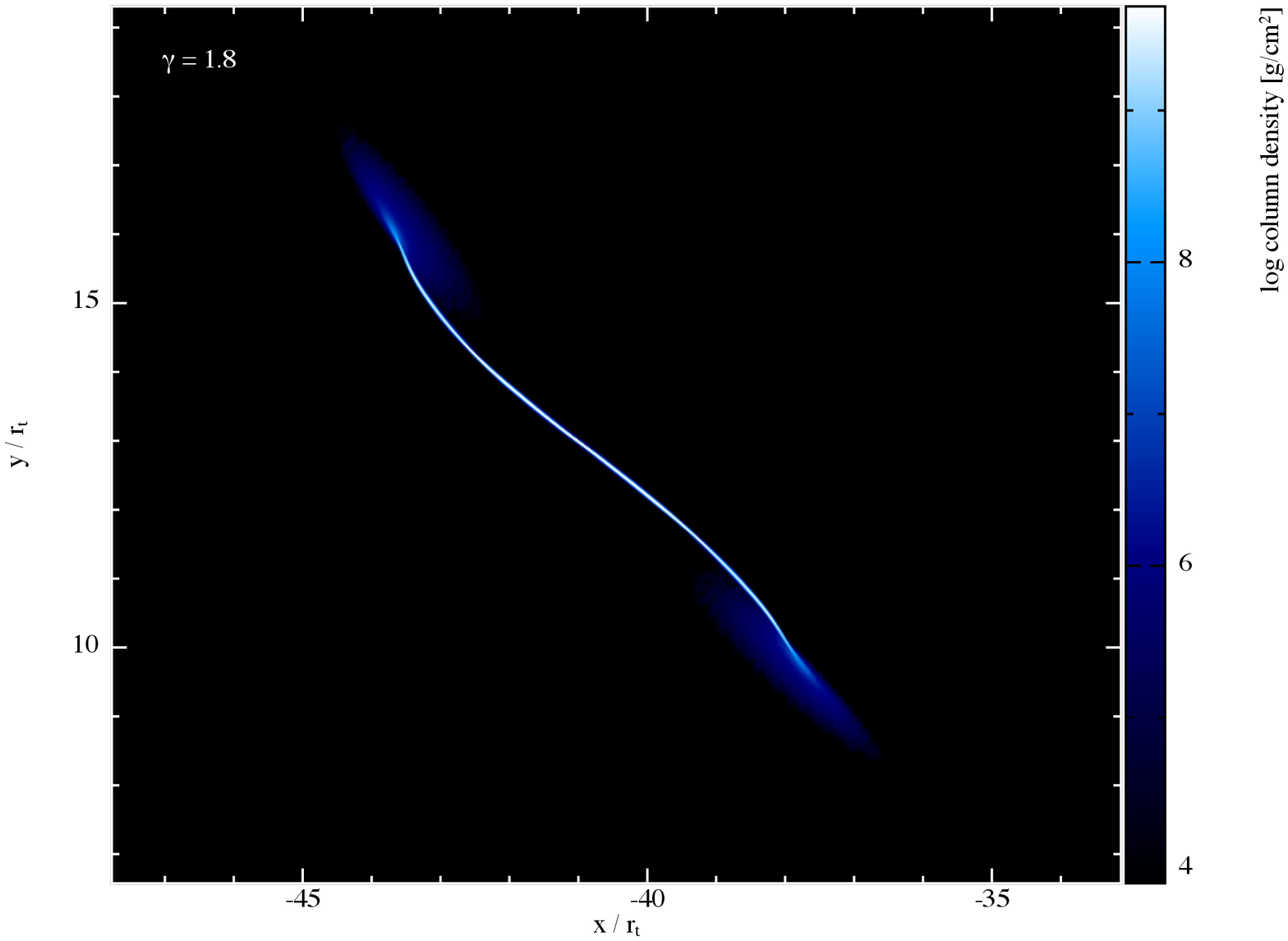} 
   \includegraphics[width=.495\textwidth, height=.43\textwidth]{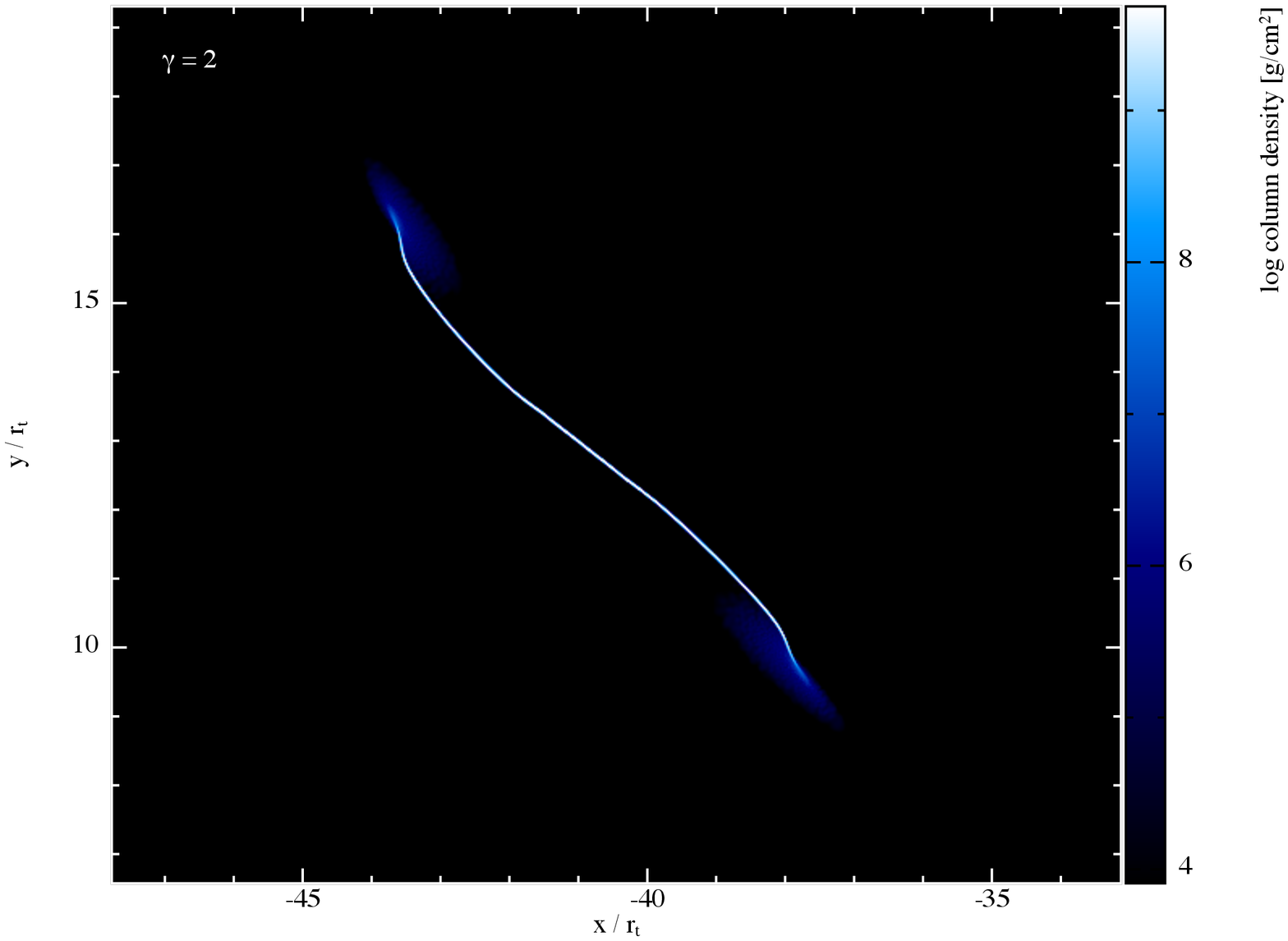} 
   \caption{The stream at a time of 2.53 days from pericenter for an adiabatic index of $\gamma = 1.5$ (top, left), $\gamma = 5/3$ (top, right), $\gamma = 1.8$ (bottom, left) and $\gamma=2$ (bottom, right). The stream thickness decreases dramatically and the fans become less pronounced as $\gamma$ increases.}
   \label{fig:2p53dplots}
\end{figure*}

Figure \ref{fig:2p53dplots} shows the disrupted stream 2.53 days after disruption for the four different adiabatic indices. In this case it is evident that a larger adiabatic index corresponds to a thinner, denser stream. This result may seem counterintuitive, as one might expect the higher-density core of the lower-$\gamma$ polytropes to result in a denser stream. However, if one assumes that pressure and self-gravity are the two dominant terms controlling the width of the stream, which is a reasonable assumption because of the nature of the perpendicular pancake, then the transverse structure of the stream is governed by the equation of hydrostatic equilibrium:

\begin{equation}
\frac{1}{\rho}\frac{\partial{p}}{\partial{s}} = -\frac{\partial\phi_{sg}}{\partial{s}} \label{sg1},
\end{equation}
where $\phi_{sg}$ is the gravitational potential due to the self-gravity of the debris and $s$ is the transverse distance from the center of the stream. Furthermore, if the variation in the self-gravitational potential along the radial direction of the stream is small, which is a good approximation toward the center of the stream owing to its approximately symmetric nature and only breaks down when we approach its radial extremities, then the Poisson equation reads

\begin{equation}
\frac{1}{s}\frac{\partial}{\partial{s}}\left(s\frac{\partial\phi_{sg}}{\partial{s}}\right) = 4\pi{G}\rho.
\end{equation}
Using this equation in conjunction with equation \eqref{sg1}, we find that the equation of hydrostatic equilibrium becomes 

\begin{equation}
\frac{1}{s}\frac{\partial}{\partial{s}}\left(\frac{s}{\rho}\frac{\partial{p}}{\partial{s}}\right) = -4\pi{G}\rho \label{lane-emden}.
\end{equation}
With the polytropic equation of state $p \propto \rho^{\gamma}$, dimensional analysis of this equation shows that the cross-sectional radius of the stream varies as

\begin{equation}
H \propto \rho^{\frac{\gamma-2}{2}} \label{Hscale},
\end{equation}
where here $\rho$ is the density at the center of the stream. The precise constant of proportionality depends on the entropy of the gas and the numerical solution to equation \eqref{lane-emden}. 

It is ultimately the scaling given by equation \eqref{Hscale} that tends to outweigh the presence of a higher-density core for smaller $\gamma$. Also, if we use this expression for $H$ in equation \eqref{rhostream}, then the density along the stream varies as 

\begin{equation}
\rho = \rho_m\left(\frac{1}{\sqrt{(u')^2+u^2(\phi')^2}}\frac{\int_{\mu\xi_1}^{\xi_1}\Theta(\xi)^{n}\xi{}d\xi}{\int_0^{\xi_1}\Theta(\xi)^{n}\xi^2d\xi}\right)^{n} \label{rhoeq},
\end{equation}
where $\rho_m$ is a normalization constant, chosen such that the density equals the correct, central stellar density at the time of disruption, and $u \equiv r/r_t$.

\begin{figure*}
   \centering
    \includegraphics[width=.495\textwidth, height=.43\textwidth]{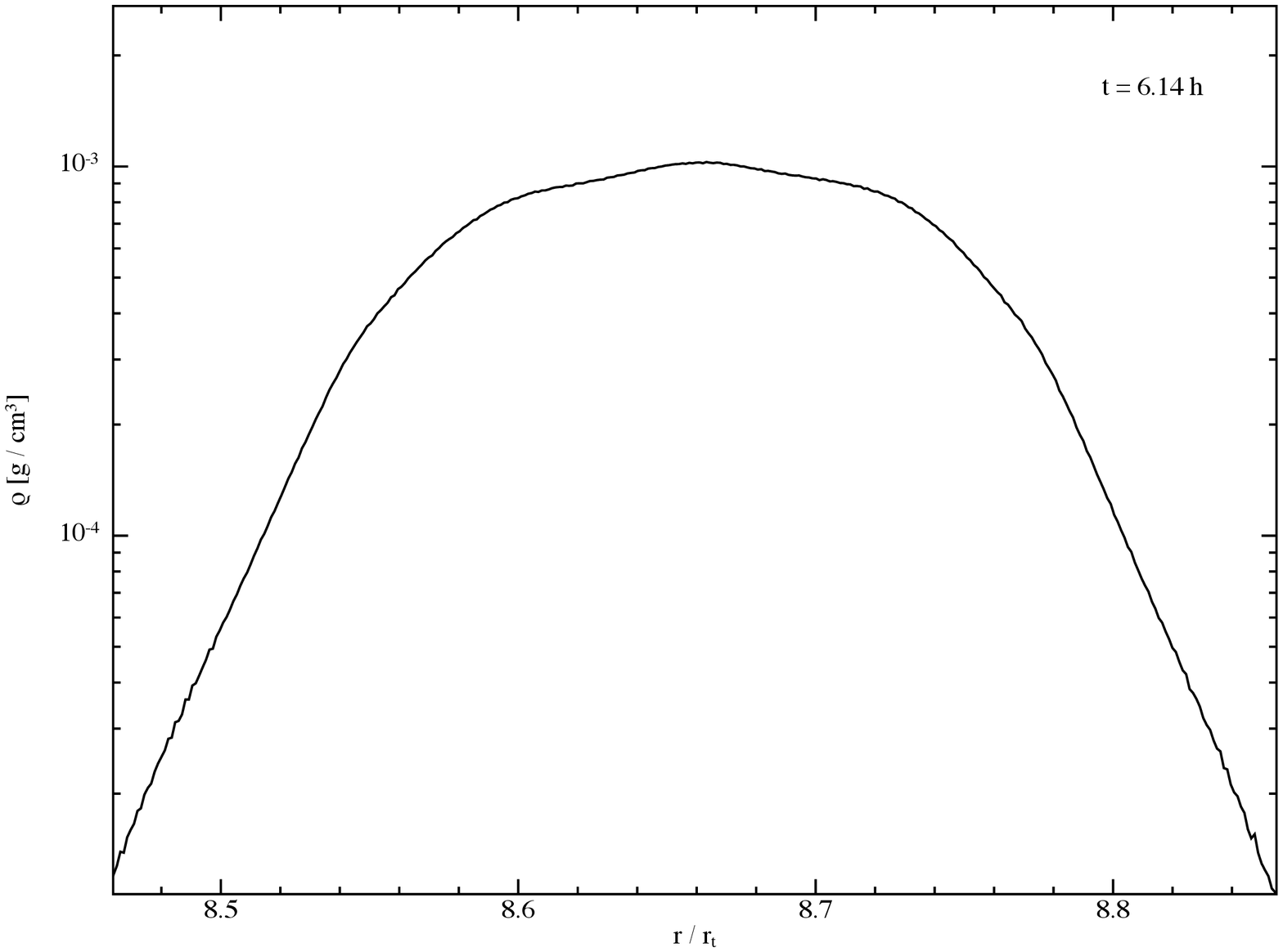}
    \includegraphics[width=.495\textwidth, height=.43\textwidth]{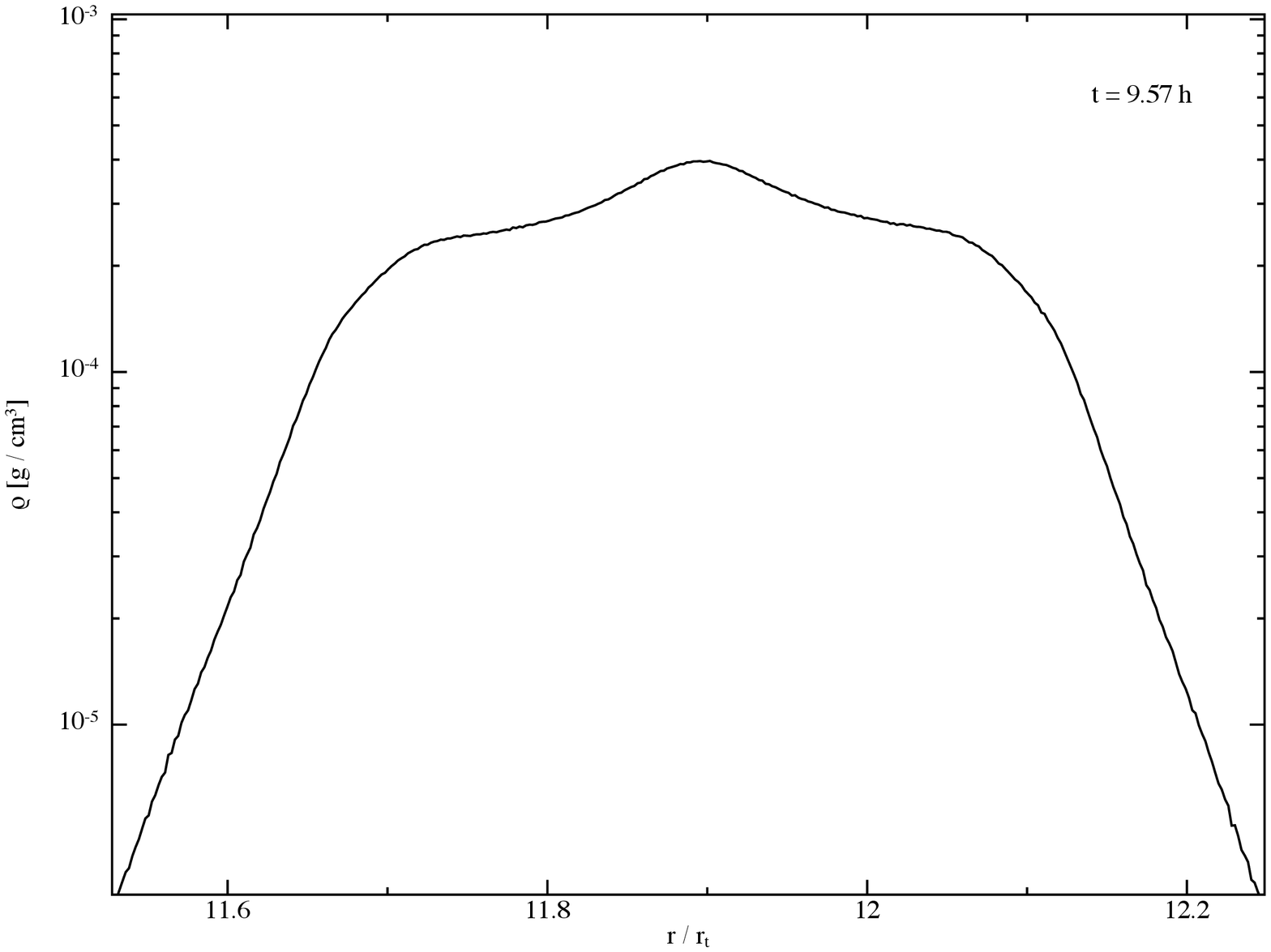}
    \includegraphics[width=.495\textwidth, height=.43\textwidth]{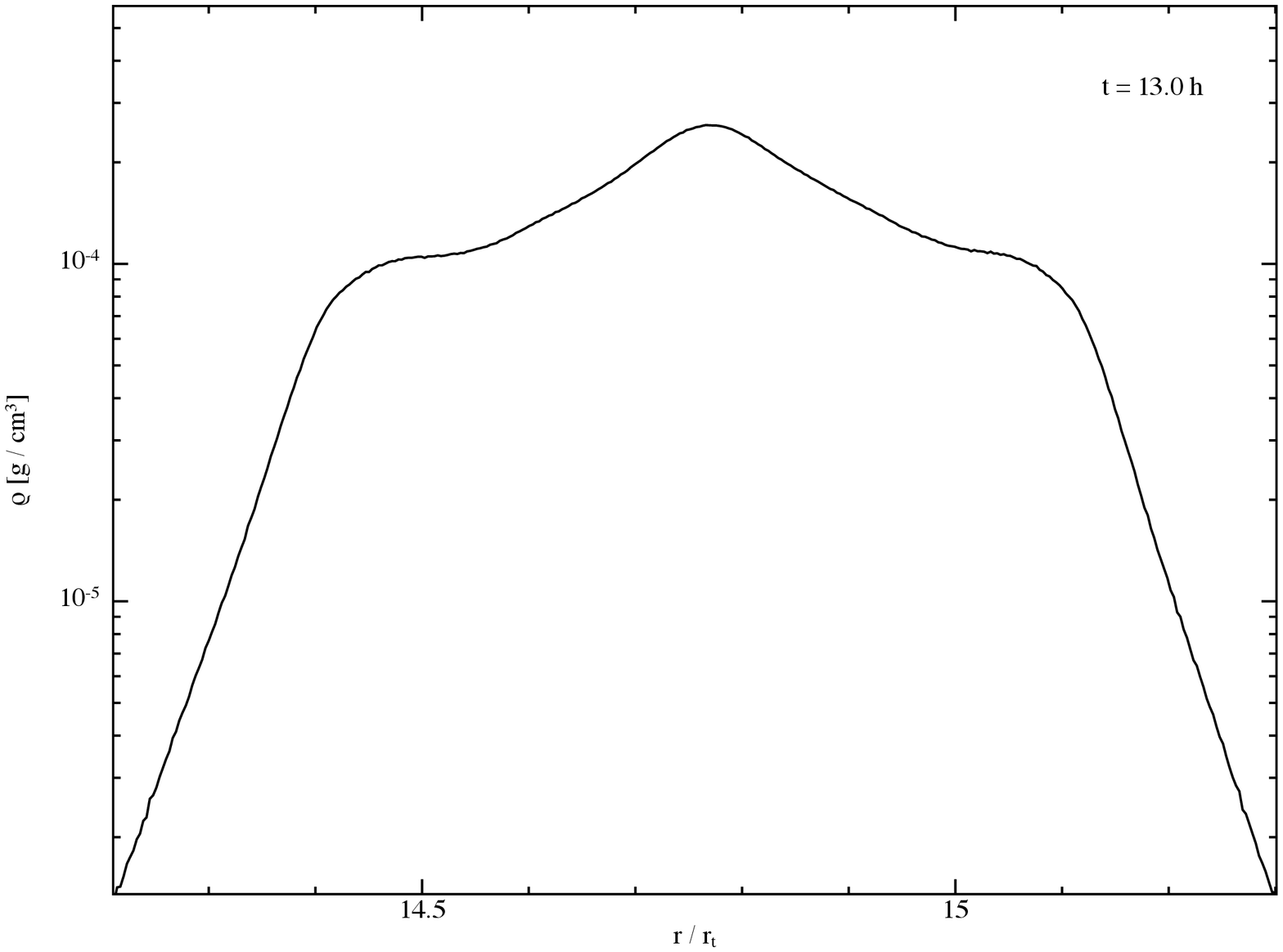}
    \includegraphics[width=.495\textwidth, height=.43\textwidth]{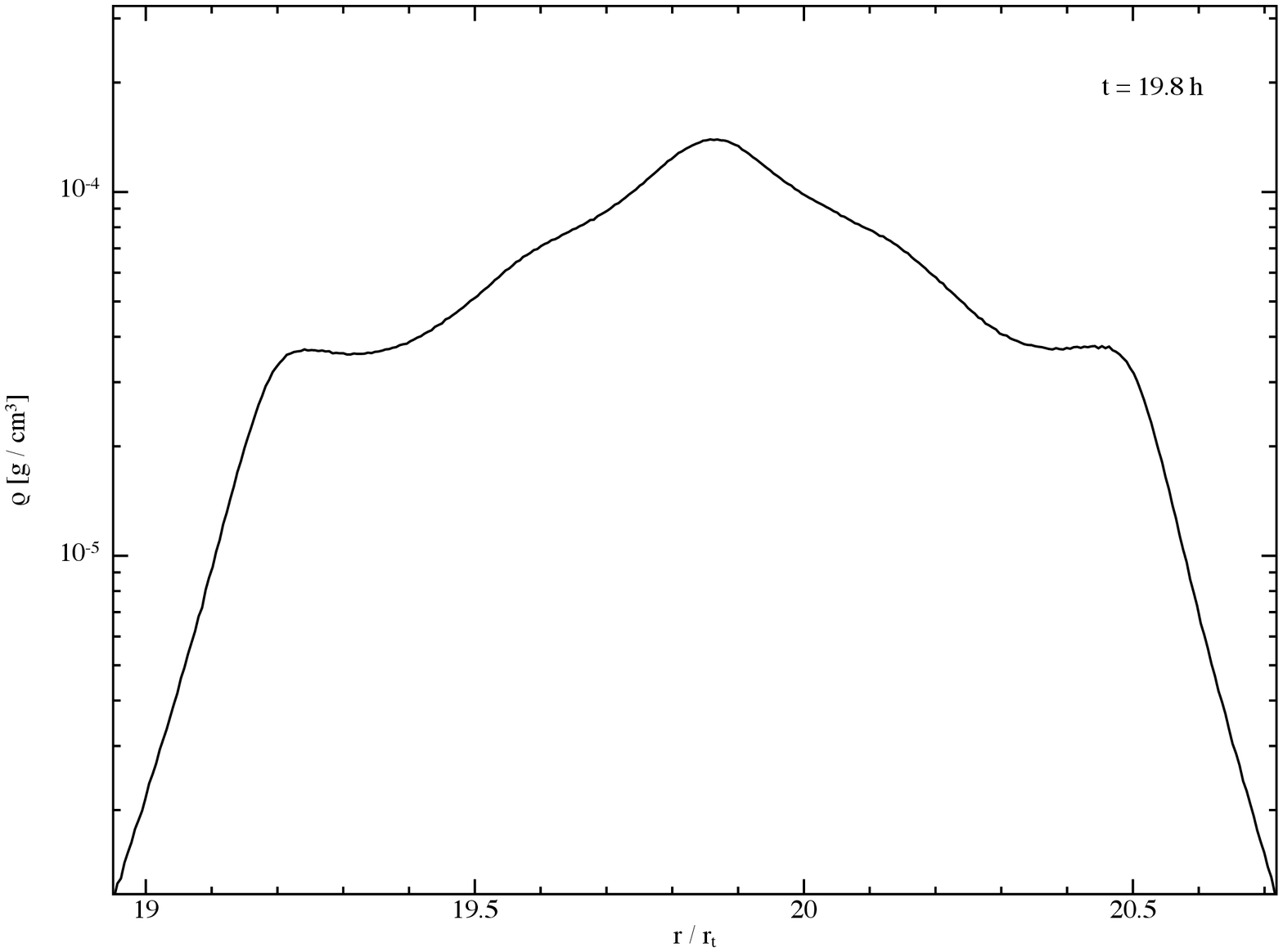}
   \caption{Four snapshots of the average {}{stream} density {}{(the average density of all particles at a given radius $r$)} as a function of $r$ for the $\gamma = 5/3$ run. Initially the density remains smooth throughout the stream; however, by about a day after the disruption, the density structure has developed a more complicated nature, consisting of a central peak that is narrower than is predicted analytically and two shoulders.}
   \label{fig:rhogamma53}
\end{figure*}

The density along the stream already exhibits a number of interesting features well before 2.53 days. To exemplify this point, Figure \ref{fig:rhogamma53} shows the {}{average radial} density {}{(i.e., the average density of all particles at a given radius $r$)} along the stream for the $\gamma = 5/3$ run at times of $t =6.14$, 9.57, 13.0, and 19.8 hours after disruption. Initially the density distribution along the curve is smooth, and matches well the distribution obtained if the original polytrope is stretched in one dimension (equation \ref{rhoeq}). However, at later times the density adopts a more intricate structure, exhibiting a sharper peak at the center of the stream and ``shoulders,'' evident from the bottom-right panel of Figure \ref{fig:rhogamma53}, that are not predicted analytically.

\begin{figure*}
   \centering
    \includegraphics[width=.495\textwidth, height=.43\textwidth]{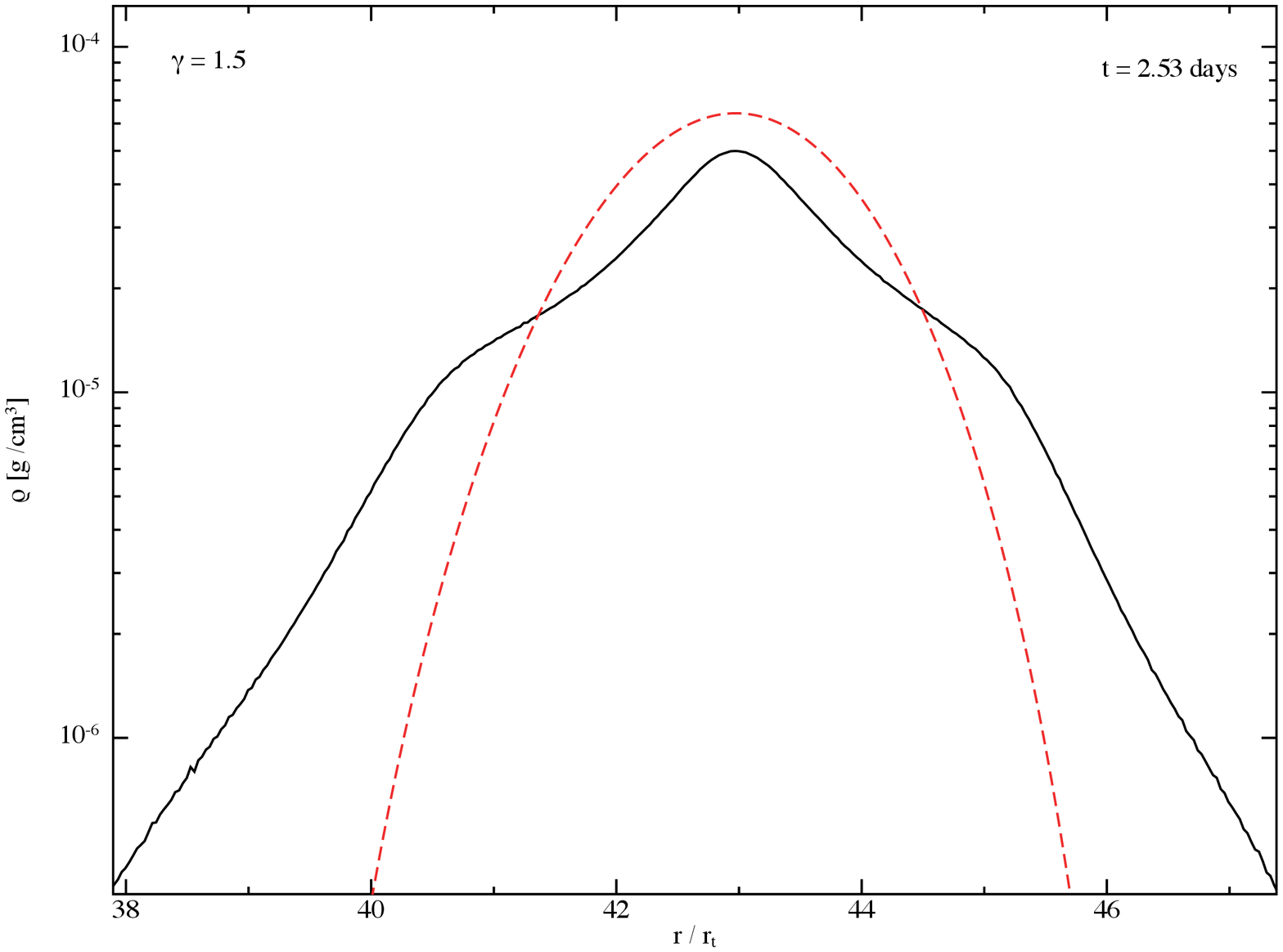}
    \includegraphics[width=.495\textwidth, height=.43\textwidth]{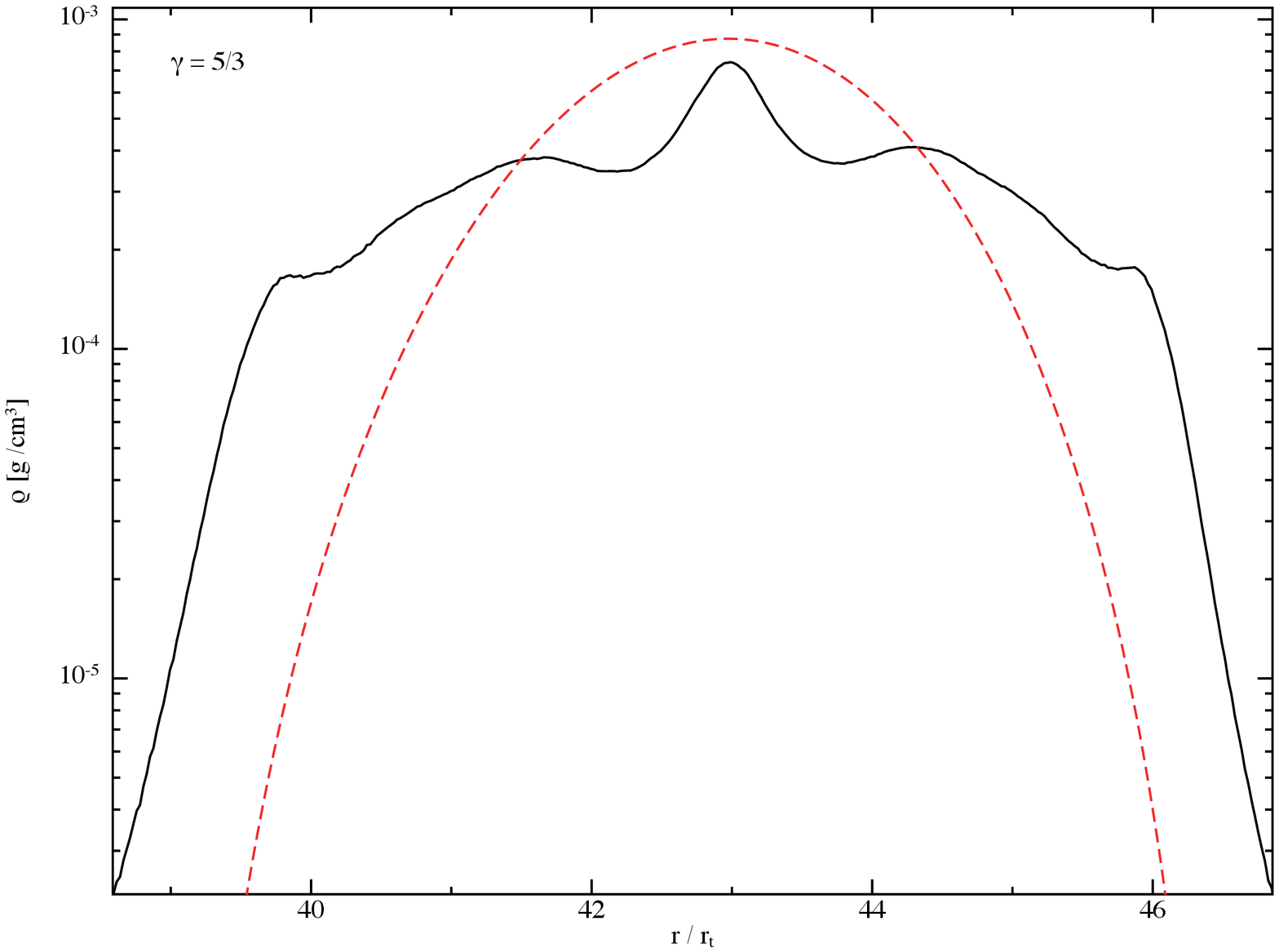}
    \includegraphics[width=.495\textwidth, height=.43\textwidth]{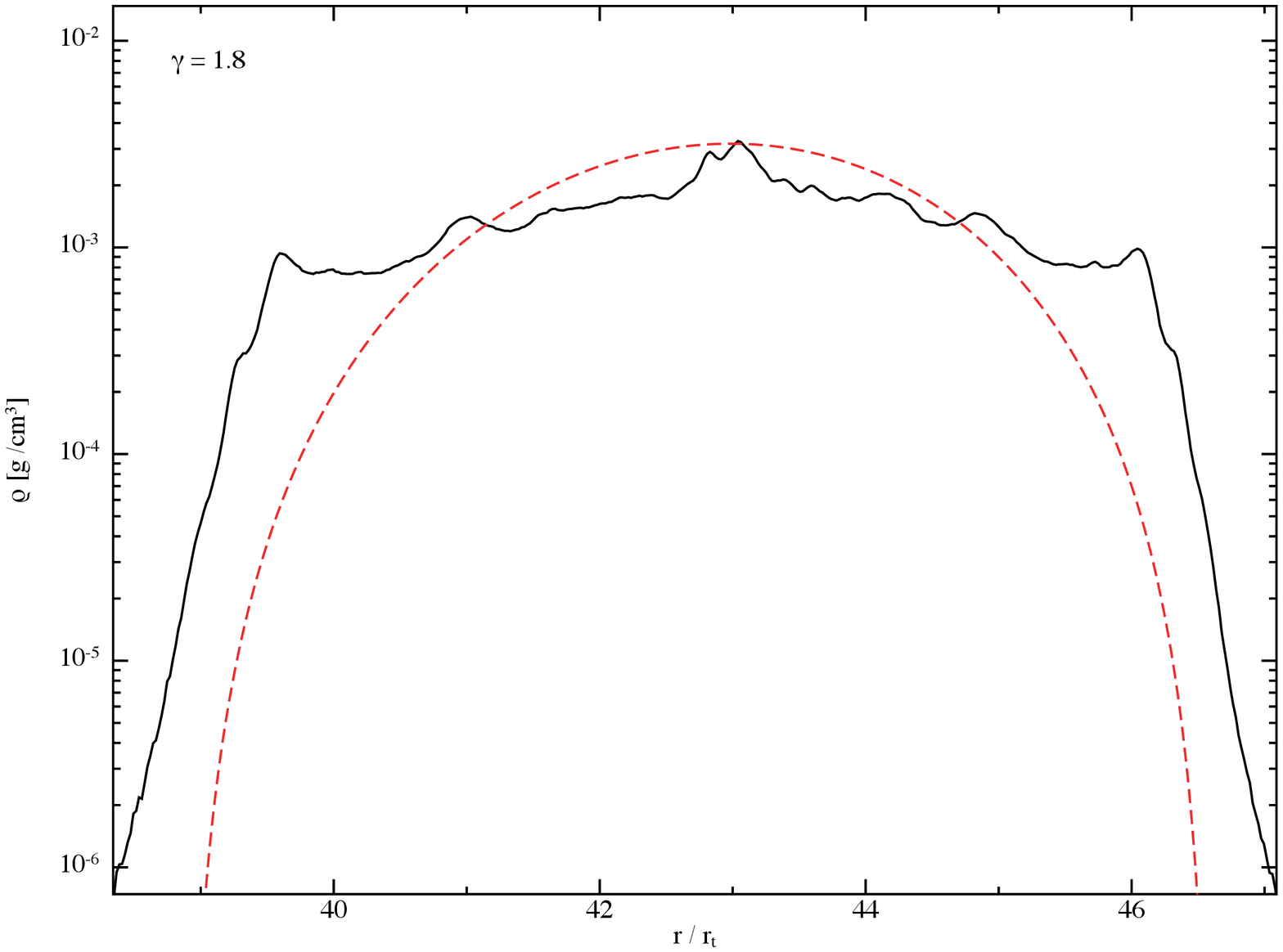}
    \includegraphics[width=.495\textwidth, height=.43\textwidth]{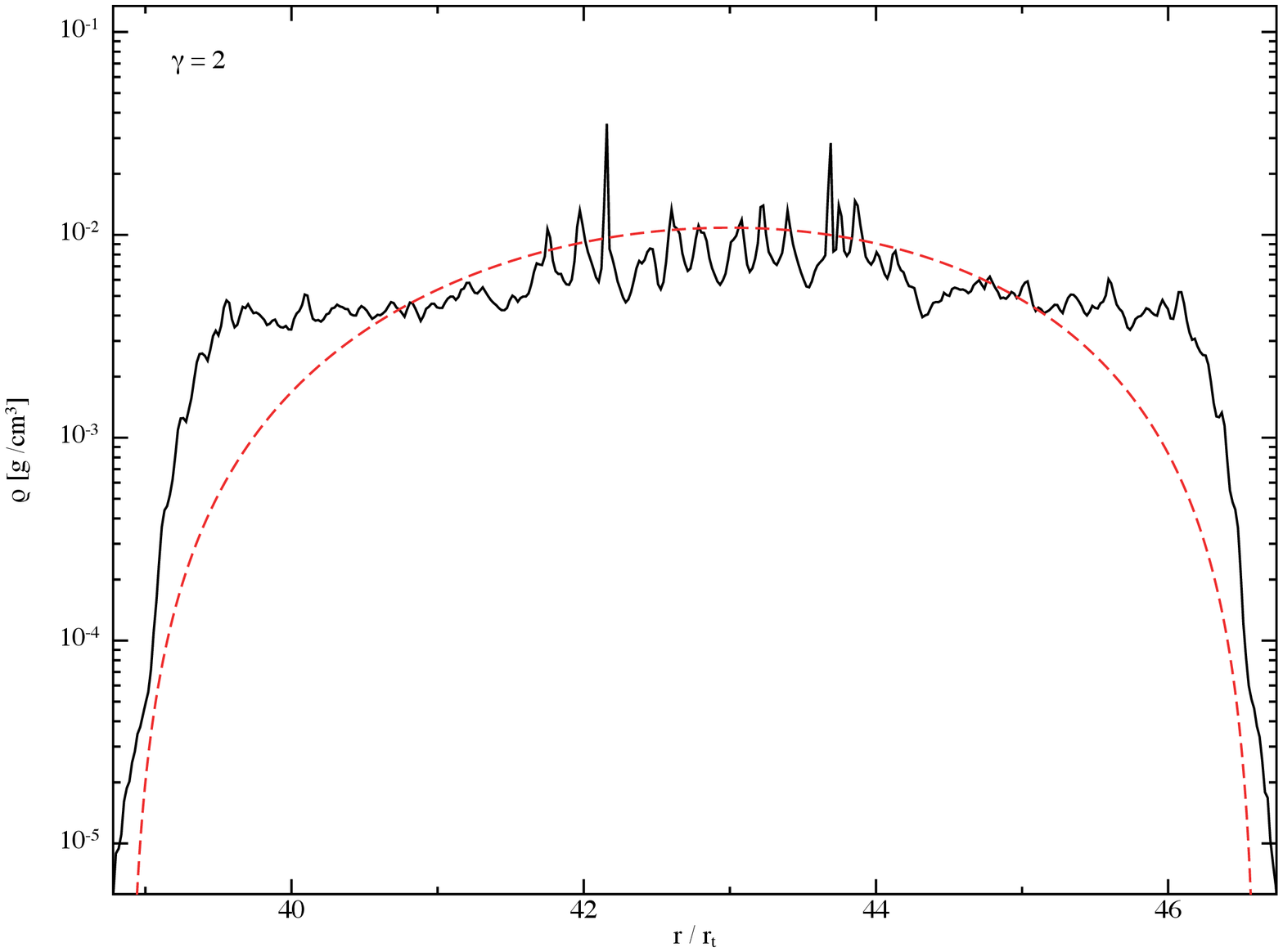}
   \caption{The {}{average} density {}{(same as Figure 6)} as a function of $r$ for $\gamma = 1.5$ (top, left panel), $\gamma = 5/3$ (top, right panel), $\gamma = 1.8$ (bottom, left panel), and $\gamma = 2$ (bottom, right panel) at a time of 2.53 days after disruption (see Figure \ref{fig:2p53dplots} for the shape of the streams at this time). The black, solid curves give the numerical solutions, while the red, dashed curves show the analytic predictions. It is apparent that larger adiabatic indices correspond to an enhanced amount of variability in the density along the stream, while a smaller adiabatic index results in more extended wings (this is also apparent from the tidal fans in the edges of the streams in Figure \ref{fig:2p53dplots}.)}
   \label{fig:rhoallgammas}
\end{figure*}

Figure \ref{fig:rhoallgammas} shows the {}{average} density along the stream for the four different adiabatic indices at 2.53 days after disruption (the black curves are the numerical solutions, while the red, dashed curves give the analytic estimate that results from equation \ref{rhoeq}). This Figure demonstrates that the small-scale density fluctuations that develop along the stream at later times are intensified for larger $\gamma$. It is evident that lower adiabatic indices show relatively smooth variations in the density, and retain an approximately symmetric structure about the center of the stream. For larger polytropic indices, however, the scale at which perturbations develop along the stream decreases and the perturbations themselves become more erratic in amplitude and position. It is also clear that a smaller adiabatic index results in more material at smaller and larger radii than would be predicted analytically, and these ``fans'' are also apparent from Figure \ref{fig:2p53dplots}. This results from the fact that polytropes with smaller $\gamma$ have lower-density envelopes, those envelopes being more easily stripped at early times.

\begin{figure}
   \centering
   \includegraphics[width=3.4in]{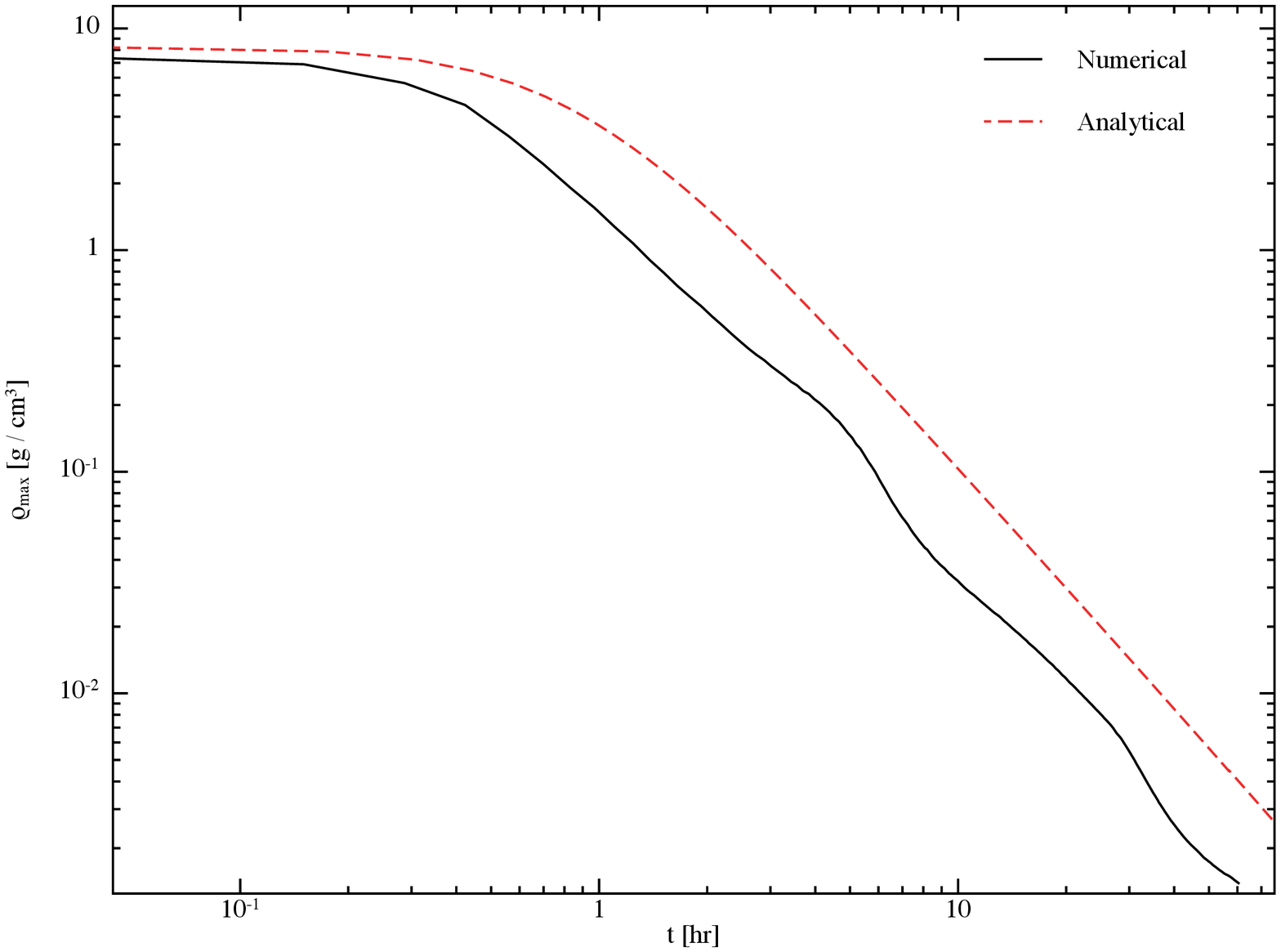} 
   \caption{The maximum density along the stream as a function of time with $\gamma = 5/3$ ($n=1.5$); the numerical solution is given by the black, solid curve, and the analytical solution (equation \ref{rhoeq}) is given by the red, dashed curve. A time of zero here corresponds to the time at which the star reaches the tidal radius. The time at which the numerically-obtained density starts to decrease is slightly earlier than the analytic one, suggesting that the time at which the star is ``disrupted'' is actually pre-periapsis. The first bump in the numerical solution, which occurs after a couple of hours, indicates where the pancake starts to augment the maximum density. At late times, both solutions follow the approximate power-law decline $\rho \propto t^{-1.8}$. }
   \label{fig:rhomaxanalytic}
\end{figure}

\begin{figure}
   \centering
   \includegraphics[width=3.4in]{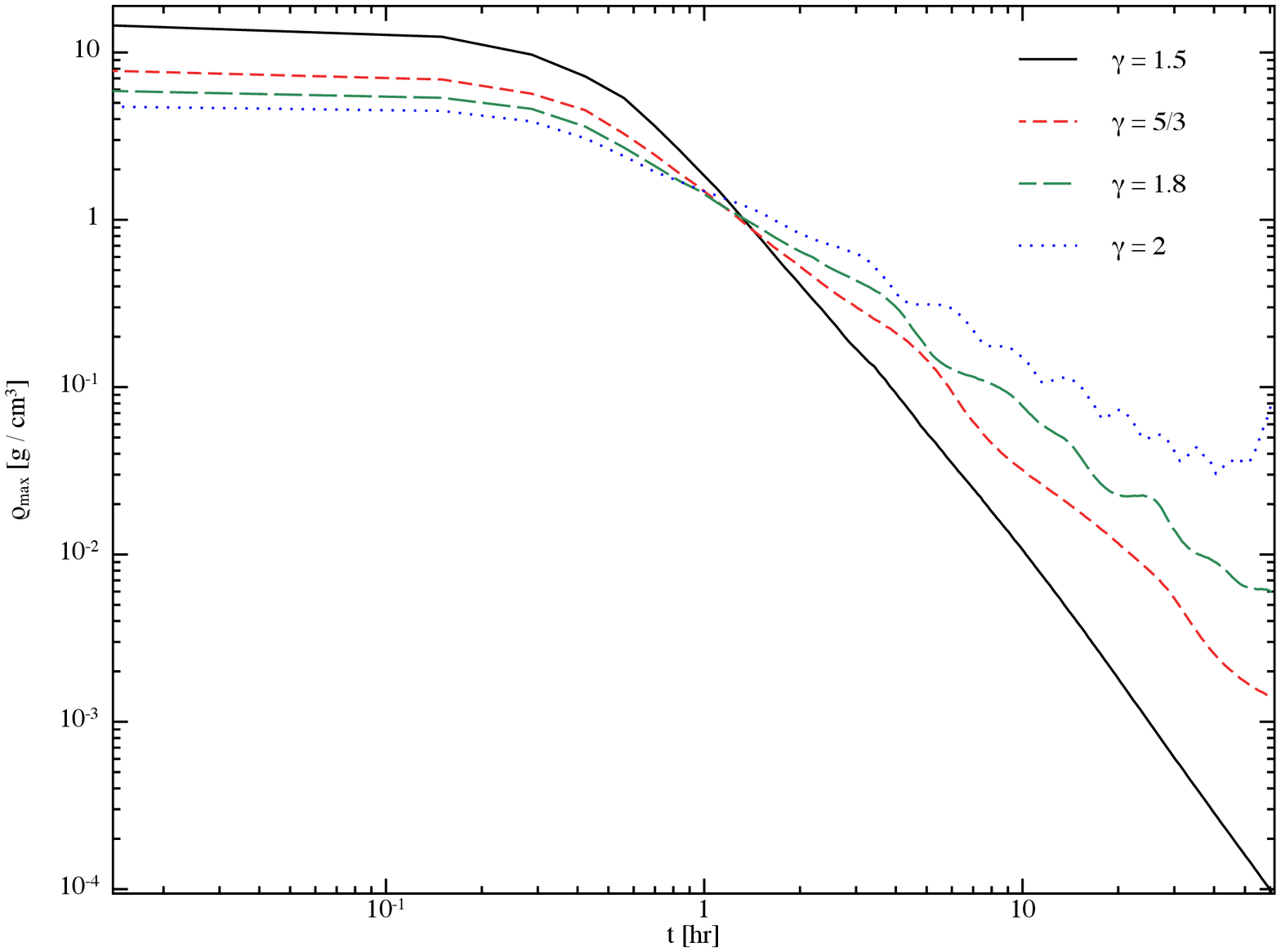} 
   \caption{The maximum density as a function of time for $\gamma = 1.5$ (black, solid curve), $\gamma = 5/3$ (red, dashed curve), $\gamma = 1.8$ (green, long-dashed curve), and $\gamma = 2$ (blue, dotted curve). It is apparent that the initial perturbation induced by the pancake is induced sooner for larger $\gamma$, and the oscillation timescale of the perturbation is shorter for larger $\gamma$.}
   \label{fig:rhomaxvt_allgammas}
\end{figure}

To determine when the density of the debris stream starts to exhibit the anomalous, small-scale structure that is apparent in Figures \ref{fig:rhogamma53} and \ref{fig:rhoallgammas}, Figure \ref{fig:rhomaxanalytic} shows the maximum density along the stream as a function of time; the black, solid curve indicates the numerical solution, while the red, dashed curve gives the analytic prediction (equation \ref{rhoeq}). Aside from slightly over-predicting its magnitude, the analytic solution matches the numerical one well, which shows that the stream approximately maintains hydrostatic balance {}{for all times during the disruption} in the transverse direction. Note that this result contrasts the findings of \citet{koc94}, who assumed that the stream was in free expansion {}{until three dynamical times post-disruption, which is roughly 1.5 hours for the disruption of a solar-type star by a $10^6M_{\astrosun}$ hole} (however, {}{the assumption of free expansion may hold in the limit of $\beta \gg 1$}). This plot also demonstrates that the first perturbation to the density appears at a couple hours after disruption, resulting in a ``ripple'' that over- and under-estimates the average value. The perturbations induced on the stream therefore behave as compression-rarefaction waves.

Figure \ref{fig:rhomaxvt_allgammas} shows the maximum density along the stream for the four different adiabatic indices. It is evident that the first bump in the density occurs slightly sooner for larger $\gamma$, appearing at around an hour for $\gamma = 2$, and that the temporal frequency of the perturbations increases as $\gamma$ increases. The average maximum density also falls off as a power-law for late times, which agrees with the analytic prediction (Figure \ref{fig:rhomaxanalytic}), with the power-law index being shallower for larger $\gamma$. In particular, if we set $\rho_{max} \propto t^{-m_{\gamma}}$, we find $m_{1.5} = 2.4 $, $m_{5/3} \simeq 1.8$, $m_{1.8} \simeq 1.5$, and $m_2 \simeq1.2$. 

\section{Discussion}
We saw in the previous subsection that the impulse approximation -- assuming that the star retains its spherical, undisturbed structure until it reaches the tidal radius -- does a reasonable job of fitting the numerically-obtained density profile of the tidally-disrupted debris stream when $\beta = r_t/r_p = 1$ (Figure \ref{fig:rhoallgammas}). This agreement demonstrates that the stream width is set by hydrostatic balance, while the length is determined by the radial positions of the gas parcels orbiting in the potential of the black hole. However, at times corresponding to a few hours after disruption, the density profile begins to exhibit anomalous, small-scale structure that is not predicted analytically, with important ramifications for the late-time evolution of the stream (Figures \ref{fig:rhoallgammas} --\ref{fig:rhomaxvt_allgammas}).

This behavior was also noted by \citet{lod09}, who commented on the existence of the shoulders present in the density profile (see their Figure 7; they were interested in the behavior of $dm/d\epsilon \propto \rho{H^2}$, the distribution of mass in energy space, as this yields information about the fallback rate). Since they renormalized their specific energy distribution to match the peak, they did not notice the sharper structure exhibited by the density in the central portion of the stream. They argued that these shoulders arose from shock compression within the stream.

However, we find it unlikely that shocks alone can account for these anomalous features. For one, shocks occur primarily in the outermost regions of the envelope at the time of disruption. The majority of the material involved in the shocks is therefore confined to the tidal tails of the debris stream (the fans at the edges of the streams in Figure \ref{fig:2p53dplots}; see Figure 8 of \citealt{lod09}), comprising only a small fraction of the total amount of mass contained in the stream. However, the perturbations occur throughout the majority of the stream, affecting a much larger fraction of the material. The time at which the fluctuations begin to appear is also hours after the disruption, well after the shocks that occur at pericenter. Furthermore, we have run additional simulations that include shock heating; in these cases, the density profiles we find are nearly identical to those presented here, indicating that the amount of material that shocks significantly is small.

On the contrary, we find that a more reasonable origin for the anomalous structure present in the numerical solutions is the \ed{combination of self-gravity} and the ``perpendicular pancake'' discussed in Section 2.2 -- where in-plane compression of the star causes the front and back edges of the star to converge to a one-dimensional line, or caustic (see Figure \ref{fig:pancakes}). This interpretation is supported by the temporal coincidence of the ripples present in Figure \ref{fig:rhoallgammas} and the analytic prediction of when the caustic arises, both occurring on the order of hours after disruption. We also note that the majority of the stream, not just the central maximum, seems to be undergoing an increase in density when the first perturbation occurs. This can be seen from Figure \ref{fig:rhomaxanalytic}, which shows that the first increase in the density for the $\gamma = 5/3$ run starts to appear around a few hours after disruption. However, the top, left panel of Figure \ref{fig:rhogamma53} shows that at a time of roughly six hours after disruption, long after the first perturbation has started to augment the maximum in the density, the entire stream still retains a smooth density distribution that is well-matched by the analytic prediction. Indeed, the sharper peak and shoulders do not seem to appear until around 10 hours after disruption, which is the top, right-hand panel of Figure \ref{fig:rhogamma53}. This indicates that the first increase in the maximum density is occurring over the \emph{entire} stream, not just in the central region where the maximum occurs (for which Figure \ref{fig:rhomaxanalytic} applies), and the density everywhere is being incremented by the same factor. 

In further support of the interpretation that the caustic \ed{occurs in the simulations and enhances the density perturbations}, recall that the existence of the caustic is ultimately related to the initial conditions at the time of disruption: because every gas parcel is moving with the center of mass of the star, the parcels comprising the back edge of the star have not yet reached their periapses, while those comprising the front have already passed through theirs. This configuration then causes the back to accelerate and the front to decelerate, resulting in their eventual merger. In a realistic TDE, the star does not retain perfect spherical symmetry all the way until the tidal radius (Figure \ref{fig:todplots}). In particular, the less dense, outer regions of the envelope will be stripped earlier, causing them to violate the condition that they move with the center of mass. The denser, central regions, however, may better retain their unperturbed structure, resulting in a pancake that occurs mainly in the center of the stream.

\begin{figure*}
   \centering
   \includegraphics[width=0.32\textwidth]{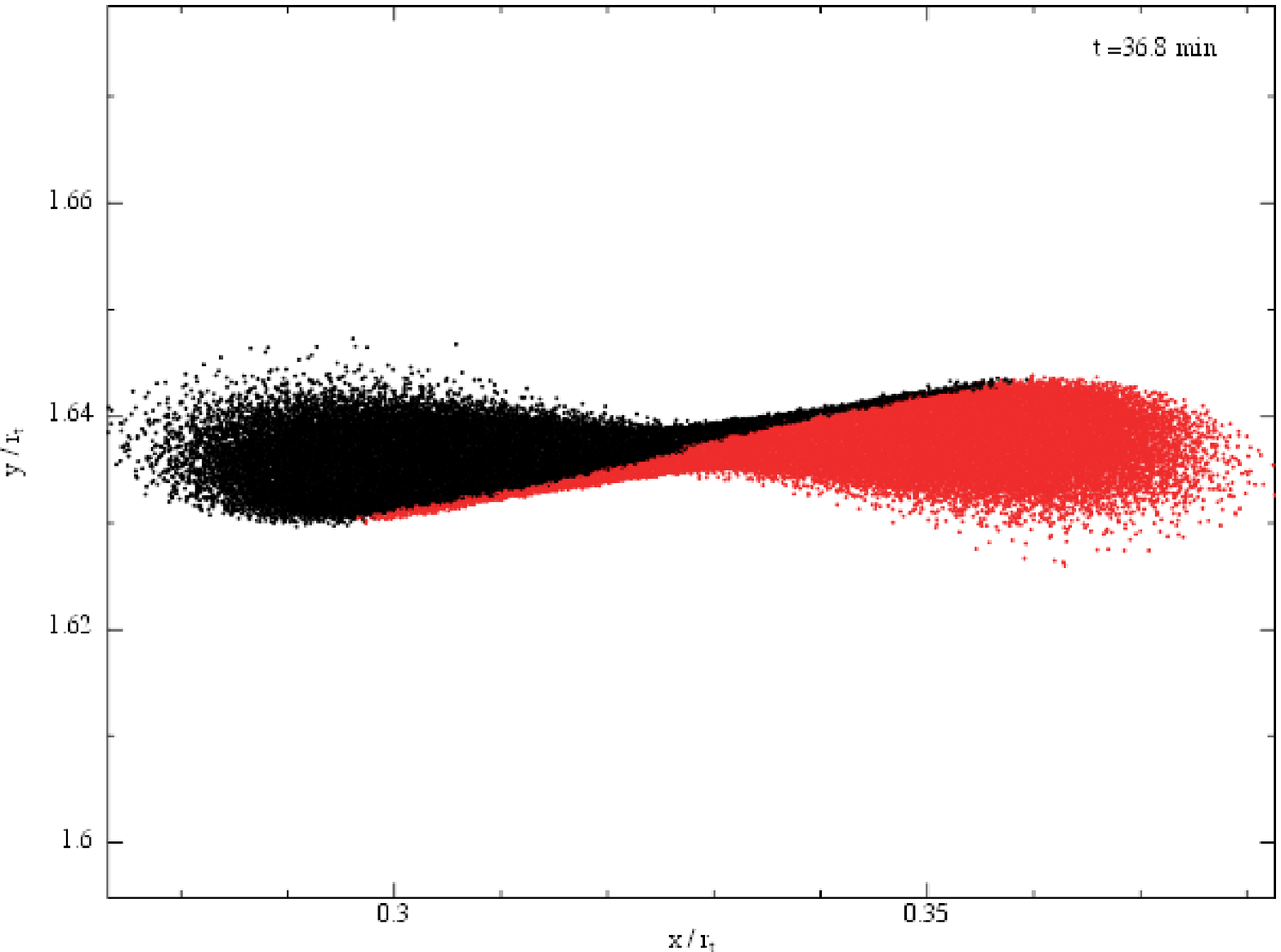} 
   \includegraphics[width=0.32\textwidth]{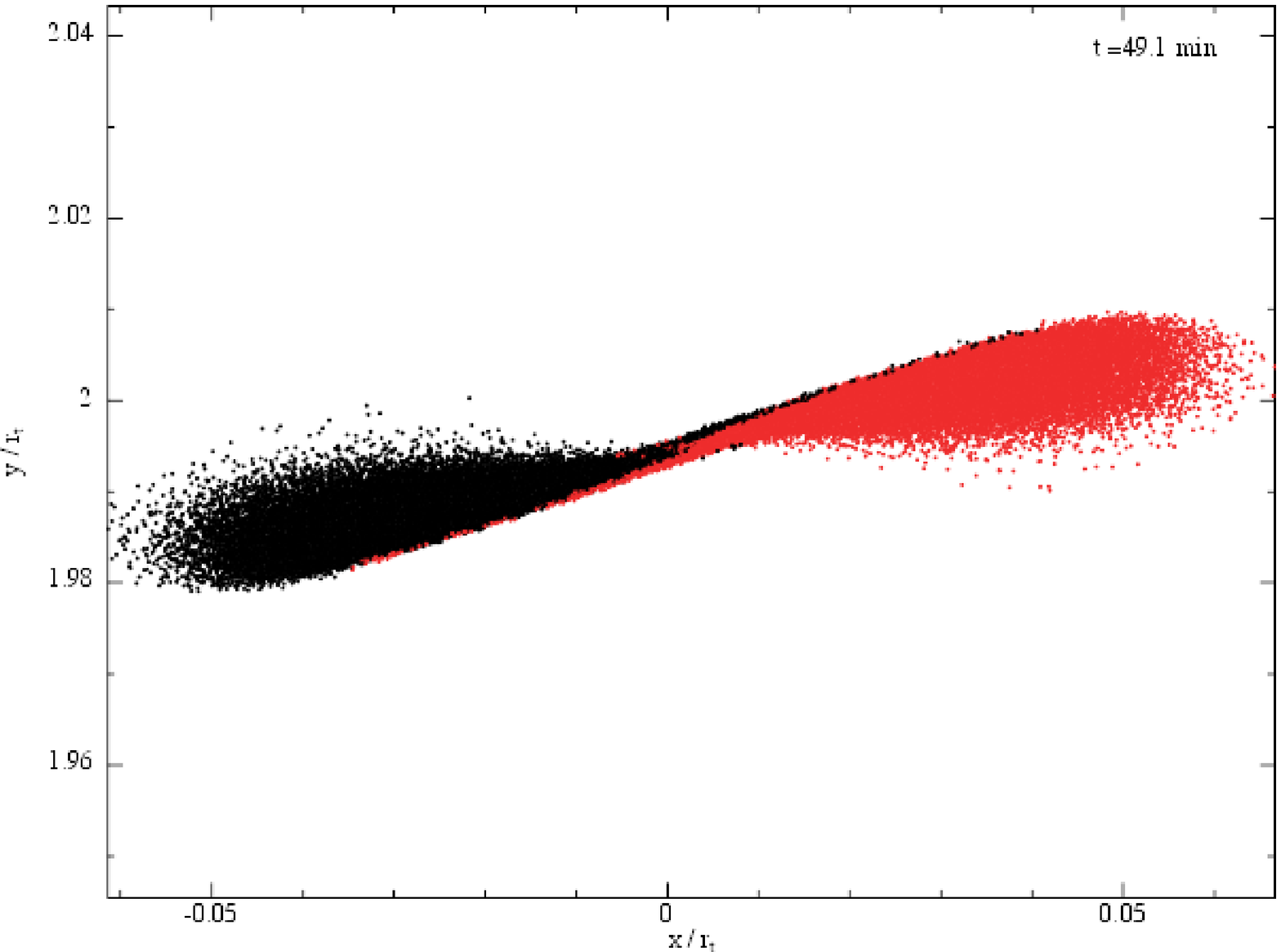} 
   \includegraphics[width=0.32\textwidth]{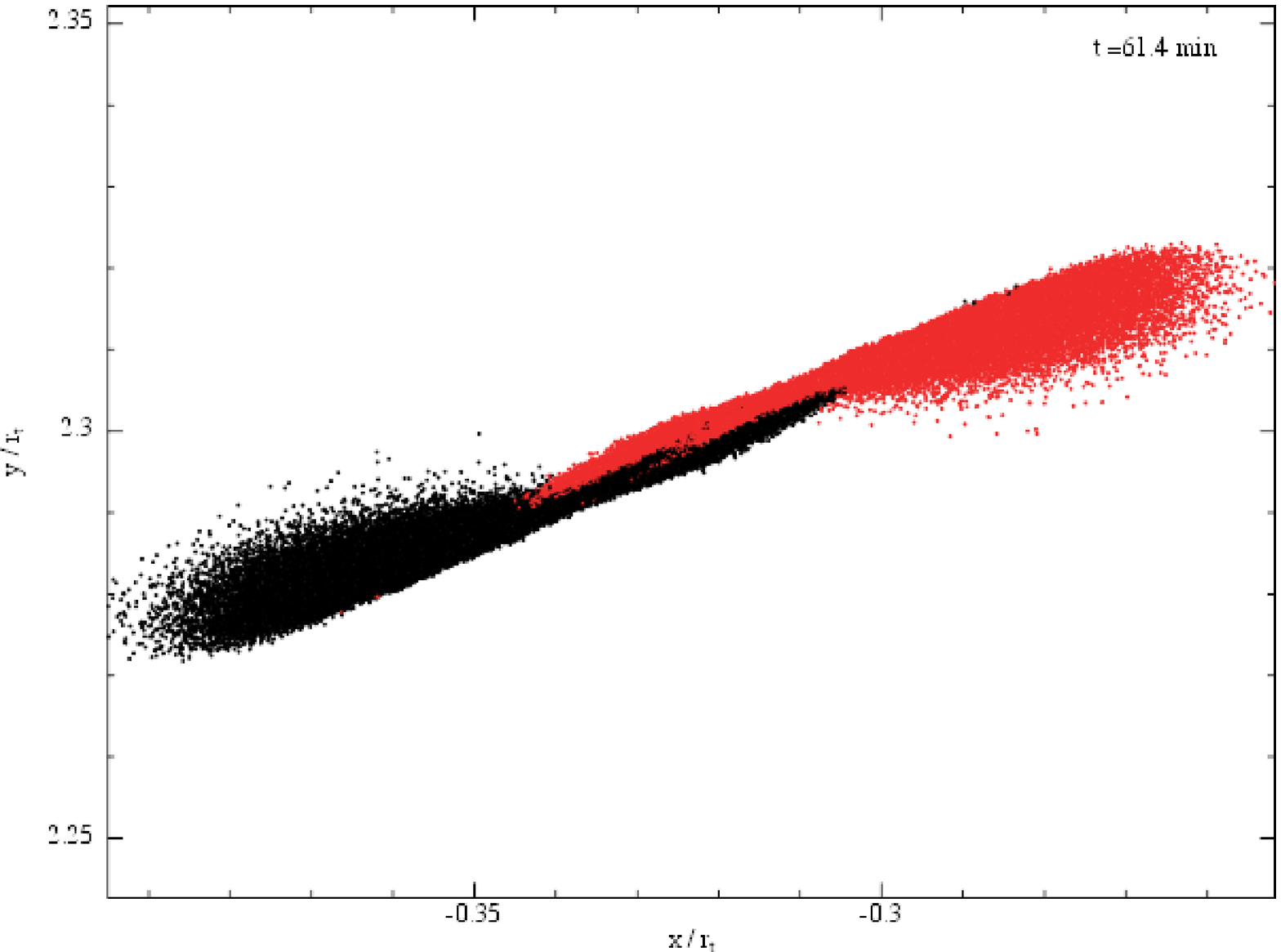} 
   \caption{The particle distributions from an $N$-body simulation, where the initial conditions were taken from the $\gamma = 5/3$ run at periapsis, {}{at 37 minutes (left panel),  50 minutes (middle panel), and 62 minutes (right panel) post-disruption. The red particles comprised the back of the star at the time of disruption, while the black constituted the front of the star. This Figure demonstrates that a caustic -- where the front and back of the stream merge to form an infinitely thin line -- still occurs in the presence of realistic initial conditions. However, as mentioned in the text, the outermost layers of the star that are stripped earlier (and therefore violate the frozen-in condition) do not undergo this compression; this is evidenced from the fact that the ``fans'' present in this figure undergo no distortion in terms of their color. The central panel} corresponds to the point at which the width of the stream has reached a minimum, the half-width being $H \simeq 0.1\,r_t$.}
   \label{fig:nbodyplot}
\end{figure*}

To test this hypothesis, we ran a simulation in which we took the output of the $\gamma=5/3$ {\sc phantom} run when the star reached pericenter and evolved the particles solely in the gravitational field of the hole, neglecting pressure and self-gravity. Figure \ref{fig:nbodyplot} shows the distribution of particles at a time of {}{37 minutes (left panel),  50 minutes (middle panel), and 62 minutes (right panel) post-disruption, the red particles being those that composed the back of the star at pericenter, the black particles the front. This Figure shows that, at roughly an hour after disruption, the front and back edges of the stream switch places, with the point of maximum compression occurring in the middle panel.} Specifically, the half-width of the center of the stream at 50 minutes is roughly $H \simeq 0.1 r_t$, which is only five times the value when the star is at pericenter. This Figure confirms that the caustic still exists with realistic initial conditions. However, as predicted, the fans at the edge of the stream do not undergo a similar amount of compression {}{and retain their original colors}, which is due to the fact that they were not moving with the center of mass at the time of disruption (i.e., they were already stripped from the star; this is also supported by the fact that the fans extend farther in radially than the analytic solutions predict, which is apparent in Figure \ref{fig:rhoallgammas}).

The preceding arguments illustrate that it is likely the caustic discussed in Section 2.2 that \ed{augments the importance of self-gravity and} generates the density fluctuations in the stream. Interestingly, Figures \ref{fig:rhomaxanalytic} and \ref{fig:rhomaxvt_allgammas} show that this perpendicular pancake does not simply increase the density, but instead generates a compression-rarefaction wave. This is due to the fact that the increase in the density likewise generates an increase in the pressure, which resists the compression. Eventually, the continued squeezing of the stream results in the material being overpressured in the transverse direction, which causes the stream to ``bounce.''

The sharper peak that develops in the center of the stream arises from the self-gravity of the debris. In particular, the compression in the transverse direction augments the central density to the point where material can be drawn in gravitationally in the radial direction, which creates the more massive central peak and the two dips on either side of that peak in Figure \ref{fig:rhogamma53}. The two shoulders that develop are regions of the stream that have not been gravitationally drained of material by the central peak and are slightly denser than one would predict analytically due to the pancake. More structure develops at late times, and local maxima are imprinted due to the oscillation of the stream, ultimately due to the self-gravitating nature of the debris {}{(see also Figure 3 of \citet{koc94}, who found oscillations in the stream width and height due to pressure and self-gravity)}. The points at which the density sharply drops off are the fans present in Figure \ref{fig:2p53dplots}, and have thus not been affected by the caustic (note from Figure \ref{fig:rhoallgammas} that the rate at which the density falls off with radius in these regions parallels the analytic one, which confirms this interpretation). 

From Figure \ref{fig:rhoallgammas}, it is apparent that larger adiabatic indices result in more drastic fluctuations that are induced by the caustic. The reason for this scaling is likely two-fold, the first being that, for the same physical radius $R_*$, polytropes with larger adiabatic indices have flatter density profiles (note that this is not true in the dimensionless space spanned by $\xi$). Therefore, since the density throughout the envelope differs from that of the core only when we are near the surface of the star, polytropes with higher adiabatic indices can better retain their structure until they reach periapsis. This then results in more of the stream experiencing the effects of the caustic, which correspondingly results in a more drastic increase in the density along the majority of the stream. This is supported by Figure \ref{fig:rhoallgammas}, which shows that the shoulders extend farther from the center of the stream as $\gamma$ increases.

The second reason is that the stream is thinner for larger $\gamma$, which is evident from Figure \ref{fig:2p53dplots}. Since the equilibrium width of the stream increases as $\gamma$ decreases, the pancake is less effective in compressing the stream and correspondingly increasing the density to the point where self-gravity can amplify the perturbations. Additionally, this scaling with $H$ causes the average density of the stream to decrease less rapidly with time for larger $\gamma$ (Figure \ref{fig:rhomaxvt_allgammas}). The overdensities within the stream are therefore more dense in an absolute sense, which increases the ability of the self-gravity of the debris to counteract the tidal shear imposed by the black hole. 

\subsection{Is the pancake necessary?}
{}{Figures \ref{fig:rhogamma53} -- \ref{fig:rhomaxvt_allgammas} show that self-gravity can drastically modify the density profile of the disrupted debris stream from a TDE, causing a sharper peak near the center, small-scale fluctuations, and ``shoulders,'' all of which are not predicted analytically. These effects are long-lived, altering the structure of the debris stream for days to months post-disruption (see also section 5). In addition, we saw in Section 2 that a caustic -- where the front and the back of the stream intersect to form a two-dimensional plane -- occurs not long after the disruption of the star under the impulse approximation. Figure \ref{fig:nbodyplot} shows that, even in a realistic TDE where the frozen-in assumption does not apply, the orbits of the gas parcels near the center of the star converge to form this post-periapsis pancake. Therefore, the self-gravity of the stream is augmented by the dynamical focusing of the gas parcels in the transverse direction.}

{}{Because the numerical method treats the full complexity of the problem, including pressure, self-gravity, and the influence of the SMBH, the simulations presented here have not isolated the effects of self-gravity and the pancaking of the orbits. Is it possible that the latter is actually unimportant, with the majority of the variation in the density of the stream due solely to the self-gravity of the debris?}

{}{To answer this question, recall that the pancake arises from the fact that, under the impulse approximation, the front of the star is decelerating at the time of disruption while the back is accelerating. Equivalently, the requirement that the entire star move with the center of mass means that the gas parcels comprising the front of the star have already passed through their pericenters, while those comprising the back have not yet passed through theirs. Therefore, to avoid the caustic but still maintain a realistic distribution of specific energies (half bound, half unbound), one can simply impose that the initial velocities of the gas parcels satisfy $\dot{r}_i = 0$, $\dot\theta_i = 0$, and $r_i^2\sin^2\theta_i\dot\phi_i^2 = 2GM_h/r_t$. Thus, if the star had these (albeit contrived) initial conditions, the post-disruption evolution would be unaffected by the caustic.}

\begin{figure}
   \centering
   \includegraphics[width=3.4in]{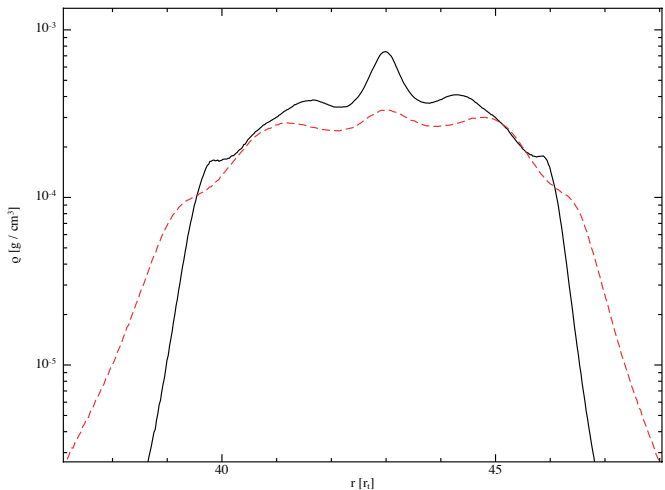} 
   \caption{{}{The average density as a function of $r$ for the unaltered, $\gamma = 5/3$ run (solid, black curve) and the run that avoids the caustic (dashed, red curve), both at a time of 2.53 days (the same time as in Figure \ref{fig:rhoallgammas}). This Figure shows that the pancake amplifies the anomalous density structures along the stream, effectively enhancing the ability of self-gravity.}}
   \label{fig:rhocomp}
\end{figure}

{}{To examine the isolated effects of self-gravity, we used the output of the {\sc{phantom}} runs when the star was at pericenter (Figure \ref{fig:todplots}) and modified the instantaneous velocities to reflect the initial conditions that avoid the caustic, i.e., we set $\dot{r}_i = 0$, $\dot\theta_i = 0$, and $r_i^2\sin^2\theta_i\dot\phi_i^2 = 2GM_h/r_t$ for all of the particles. What we generally found was that the anomalous features of the density profile were still present, i.e., shoulders still formed and a more concentrated peak developed. However, the \emph{magnitude} of each of these features was significantly reduced; in particular, the shoulders were much less pronounced, the central density peak was less sharp, and the density fluctuations were less concentrated. The overall magnitude of the density was also down by a factor of a few, and the increase in the density that occurred over the entire stream (see discussion above) was not observed in the modified runs (see Figure \ref{fig:rhocomp}, which illustrates these points). Finally, the morphology of the streams also differed, having larger widths and more extended fans in the cases where the pancake did not occur.}

{}{These tests show that, in general, the anomalous features arise from the self gravity of the debris modifying the radial density distribution throughout the stream. However, as was suggested in the previous subsection, the post-periapsis pancake is quite important for magnifying and sustaining the self-gravitating nature of the stream.} 

\section{Implications}
We have demonstrated above that a caustic, or a ``perpendicular pancake,'' \ed{augments the importance of self-gravity in the debris stream from a TDE.} In particular, we found that this pancake \ed{and self-gravity cause} density perturbations that are not predicted analytically (Figure \ref{fig:rhoallgammas}). In this section we briefly discuss several implications of our findings. 

\begin{figure*}
   \centering
   \includegraphics[width=.495\textwidth, height=.43\textwidth]{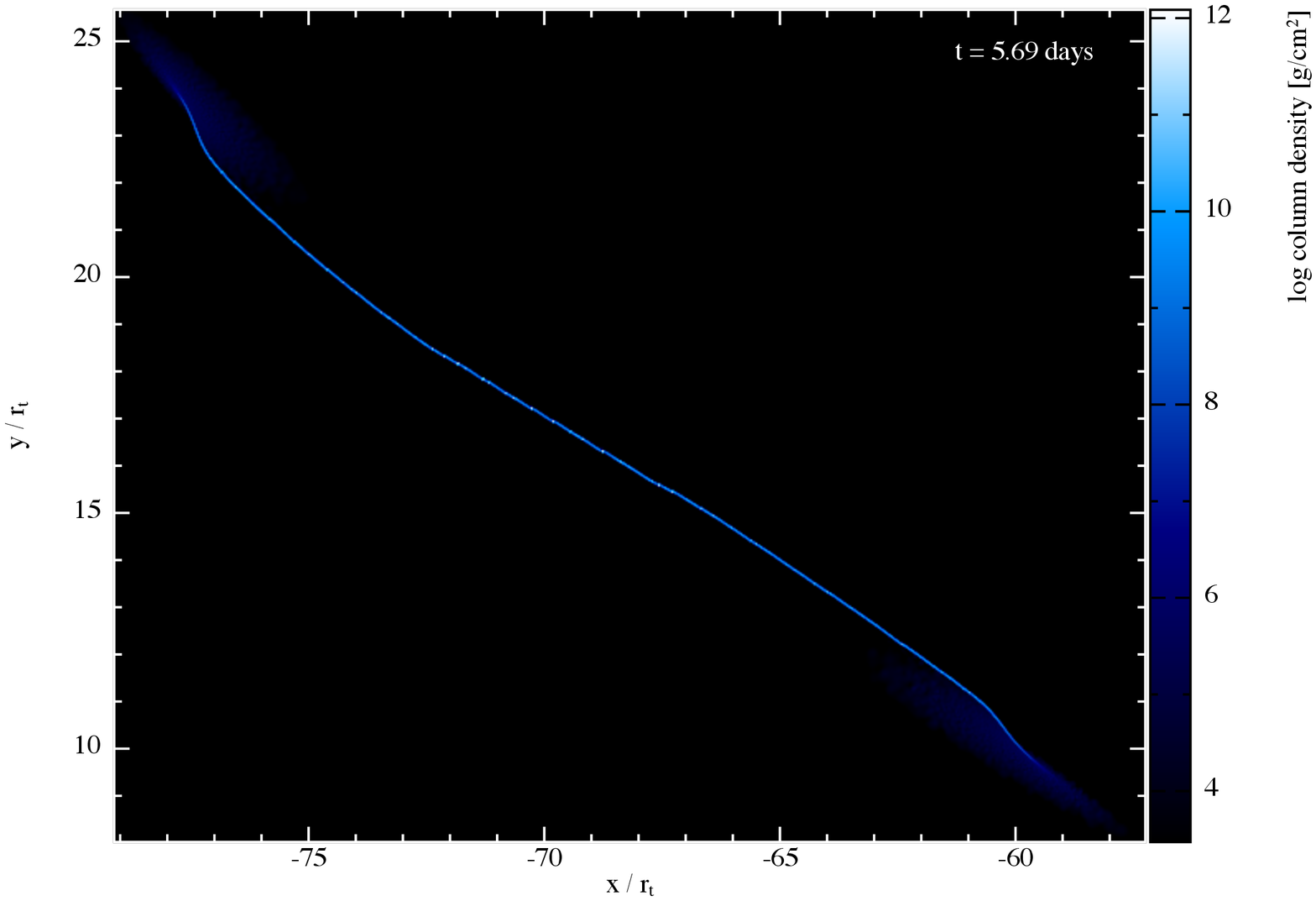} 
   \includegraphics[width=.495\textwidth, height=.43\textwidth]{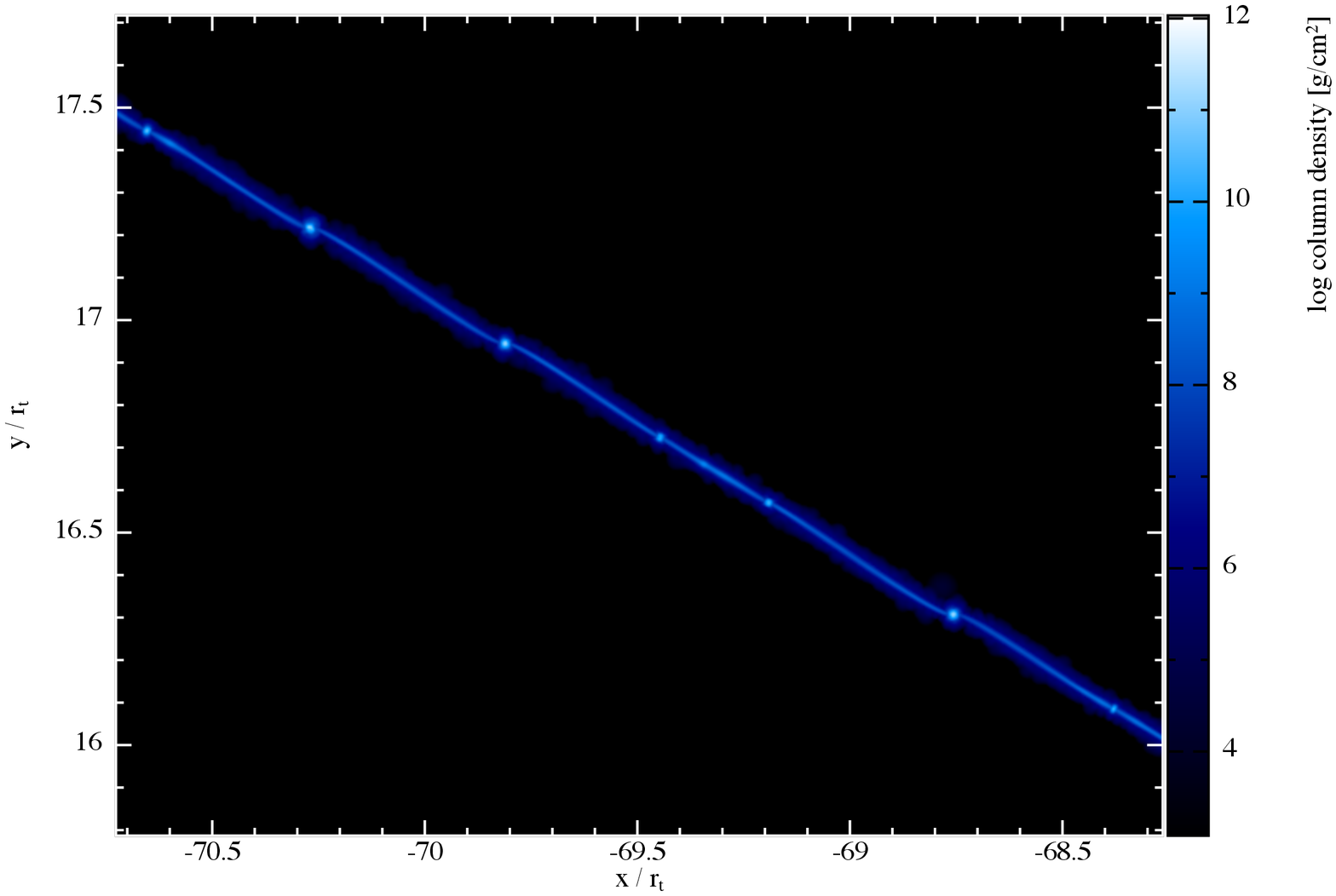} 
   \caption{The stream from the $\gamma = 2$ run (left-hand panel) and a closeup view of the stream (right-hand panel),  showing the clumps that have formed throughout the majority of the stream, both at a time of 5.69 days after disruption.}
   \label{fig:streamclumps_gamma2}
\end{figure*}

\subsection{Fragmentation}
One of the most profound implications is that these perturbations can result in the gravitational fragmentation of the stream. For $\gamma = 2$, the overdensities present in the stream at a time of 2.53 days are already self-gravitating and starting to collapse into small-scale, gravitationally-bound clumps (see Figure \ref{fig:streamclumps_gamma2}). For $\gamma = 1.8$, the stream also fragments, but not significantly until a time of a couple weeks after disruption. The $\gamma = 5/3$ run also collapses at late times, but the time at which fragmentation occurs depends on the resolution of the simulation. As commented upon by \citet{cou15}, this suggests that the stream itself is gravitationally unstable, but the perturbations induced by the pancake and self-gravity are not sufficient to drive the fragmentation. {}{This finding also suggests that the limiting adiabatic index at which fragmentation occurs is closer to $\gamma = 5/3$ than $\gamma = 2$, as indicated by previous studies of compact object mergers (\citealt{lee07}, in particular their Figure 23; see also our discussion below regarding the origin of this marginal stability)}. We have run the $\gamma = 1.5$ simulation presented here out to nearly $10$ years and have not found recollapse, suggesting that the density profile of the stream is gravitationally stable. 

In the $\gamma = 1.8$ run, the first clump forms near the center of the stream around a time of five days after disruption, with smaller-mass clumps forming at later times at distances progressively farther from the central portion of the stream. By about two months after disruption, the clump formation becomes less vigorous, and the clump masses saturate at approximately constant values with an average clump mass of $\bar{M}_{c} \simeq 0.55M_{J}$, where $M_J \simeq 9.54\times10^{-4}{M_{\astrosun}}$ is the mass of Jupiter. The maximum clump mass, however, is $M_{c,max} \simeq1.5M_{J}$, showing that the clumps span a large range in mass. 

On the other hand, the first clumps form at a time of around three days after disruption for the $\gamma = 2$ run, and instead of forming one clump in the center of the stream, between five and ten form around the same time at approximately evenly-spaced intervals along the stream (this agrees with the findings of \citealt{lee07} and other studies of the tidal tails produced during compact object mergers where very stiff equations of state were used). Fragmentation ceases with an average clump mass of $\bar{M}_{c} \simeq 2.6M_{J}$ around two weeks after disruption, and the maximum clump mass in this case is $M_{c,max} \simeq 37M_{J}$.

Since the $\gamma = 5/3$ run collapsed at late times but due to the small-scale numerical noise inherent in the simulation, additional, resolved perturbations are required to study true fragmentation in this case. This marginal instability of the stream is likely due to the fact that the maximum density in the stream drops off as $\rho \propto t^{-1.8}$ (see Figure \ref{fig:rhomaxanalytic}), whereas the ``density'' of the black hole scales as $\rho \propto 1/r^3 \propto 1/t^2$, the last proportionality resulting from the fact that the orbits of the gas parcels initially follow $r \propto t^{2/3}$. The decline in the density for the $\gamma = 5/3$ case is thus barely above that of the black hole, meaning that the stream self-gravity only outweighs the tidal shear by a small margin. Additionally, since $\rho \propto t^{-2.4}$ for $\gamma = 1.5$, we do not expect fragmentation to occur in this case, and this is consistent with what is observed from the simulation.

\begin{figure}
   \centering
   \includegraphics[width=3.4in]{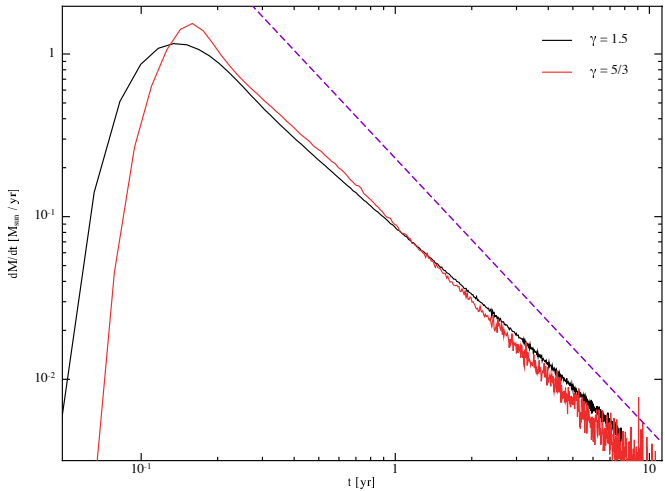} 
   \caption{The fallback rate computed for the $\gamma = 1.5$ (black curve) and $\gamma = 5/3$ (red curve) runs. The purple curve is the canonical $t^{-5/3}$ fallback rate for reference. It is apparent that the return time of the most bound material is earlier for smaller $\gamma$, which is related to the amount of distortion imparted to the star at the time of disruption. At late times, the accretion of clumps that have formed in the $\gamma = 5/3$ stream causes the fallback rate to deviate significantly from the mean (the small amount of deviation present in the $\gamma = 1.5$ run is numerical noise). }
   \label{fig:mdots}
\end{figure}

\begin{figure}
   \centering
   \includegraphics[width=3.4in]{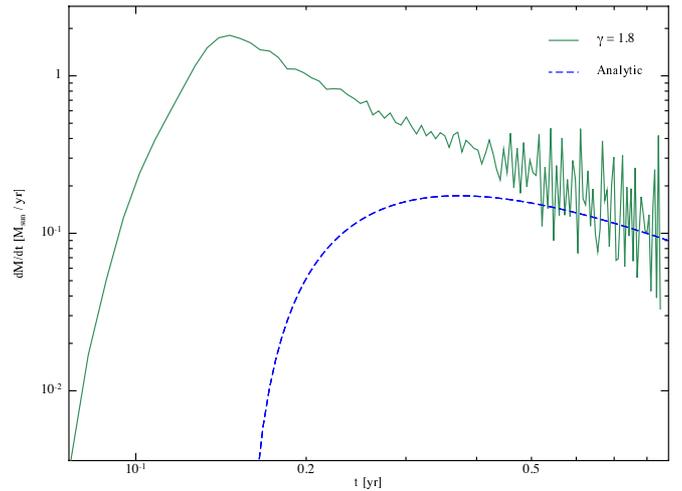} 
   \caption{The fallback rate for the $\gamma = 1.8$ run (green, solid curve) and the analytic prediction (blue, dashed curve). We see that the numerical solution is larger than the analytic one by an order of magnitude, and the fallback of bound clumps causes significant deviation from the average, $t^{-5/3}$ fallback at times greater than about 6 months from disruption. }
   \label{fig:mdot_gamma1p8}
\end{figure}

\subsection{Fallback rate features}
When one of the of these clumps returns to pericenter, the fallback rate can spike above the average, $t^{-5/3}$ decline by a significant fraction, as is apparent from Figure \ref{fig:mdots} for times greater than a few years, and from Figure \ref{fig:mdot_gamma1p8} for times greater than a few months (the small amount of scatter present for the $\gamma = 1.5$ fallback curve is due to numerical noise). If the tidally-disrupted debris has already formed an accretion disk, the interception of one of these clumps by the disk can significantly augment the accretion rate onto the hole (though this is only true during the early stages of the fallback, when the viscous time is short compared to the infall time; \citealt{can90}). 

Figure \ref{fig:mdots} shows that, shortly after reaching their peaks, the fallback rates for both the $\gamma = 1.5$ and $\gamma = 5/3$ runs fall below the canonically-assumed $t^{-5/3}$ power-law. There is then a period during which the rate is slightly shallower than the $5/3$ rate; for the 5/3 run, this latter period lasts from a few months until about a year, after which the rate resumes the $t^{-5/3}$ decay. For the $\gamma = 1.5$ run, however, the power-law is less steep than the 5/3 law after nearly ten years from the disruption. This variable fallback rate is due to the accretion of various parts of the stream: the rate drops below 5/3 when the dip between the first shoulder and the central peak of the stream (Figure \ref{fig:rhoallgammas}) is accreted. The rate then becomes shallower than the 5/3 law when the denser, central regions are accreted. This variation in the fallback rate means that observed TDEs may not follow the $t^{-5/3}$ law for much later times than previously suspected. 

Figure \ref{fig:mdot_gamma1p8} demonstrates that the peak fallback rate is significantly higher than the analytic prediction (this is also true for the $\gamma = 1.5$ and $\gamma=5/3$ runs), where the latter was calculated by using the energy-period relation, which gives $\mu(t) = (t/T)^{-2/3}$, $T = 2\pi{M_h}/(M_*\sqrt{GM_h})(R_*/2)^{3/2}$ being the period of the most tightly bound debris, and the frozen-in condition (see \citealt{cou14} for more details). This increase in the fallback rate arises from the fact that the pancake has increased the density above what would be predicted analytically, as is apparent from Figure \ref{fig:rhoallgammas}. This means that the accretion rate onto the black hole is much higher than thought previously, making it more likely that the TDE will result in a phase of super-Eddington accretion. Indeed, if we assume an efficiency of $\epsilon = 0.1$ and $\dot{M}_{acc} = \dot{M}_{fb}$ where $\dot{M}_{acc}$ is the accretion rate onto the black hole, then the peak accretion rate for the $\gamma = 1.8$ run in Figure \ref{fig:mdot_gamma1p8} corresponds to an accretion luminosity of $L_{acc} \simeq 80 L_{Edd}$, compared to the analytic estimate of $L_{acc} \simeq 8 L_{Edd}$. Since the degree to which the fallback rate is super-Eddington is inversely proportional to the black hole mass, we see that more TDEs could be accompanied by a jetted-outflow phase like that seen for \emph{Swift} J1644+57 \citep{zau11, cou14}.

\subsection{Clump fates}
If an accretion disk has not yet formed, the clumps that are bound to the black hole can return to the original pericenter distance. Since their densities will likely be lower than that of the stellar progenitor, the tidal disruption radii of the clumps will be outside the tidal radius of the original star. The returning clumps will therefore be ``redisrupted'' before reaching their pericenters, leading to complicated interactions between the streams of incoming and outgoing debris that could avoid the ``dark year for tidal disruption events'' suggested by \citet{gui15}. Also, depending on the magnitudes of general-relativistic apsidal and Lense-Thirring precession, these redisruptions may tend to isotropize the accretion process, leading to a more symmetric inflow. This symmetric inflow may then lead to super-Eddington accretion luminosities, puffing up the accretion disk and potentially leading to the production of jets \citep{cou14}. 

The clumps that form in the unbound portion of the stream will make their way out of the sphere of influence of the central SMBH and into the galaxy. In particular, if we recall that the escape velocity of the most unbound material is $v_{esc} \simeq \sqrt{2GM_*/R_*}(M_h/M_*)^{1/6}$, then we find that the unbound clumps leave the sphere of influence of the black hole on a timescale of
\begin{equation}
  t_{\rm esc} \sim 10\,\sigma_{100}^{-2} \left(\frac{M_h}{10^6M_{\odot}}\right)^{2/3} \left(\frac{M_\star}{M_\odot}\right)^{-1/6} \left(\frac{R_\star}{R_\odot}\right)^{1/2}\,~{\rm yrs},
\end{equation}
Although their long-term evolution is uncertain and depends on the specific properties of the gas (e.g., heating and cooling rates due to ionizations and recombinations), these unbound clumps could condense into planetary mass objects and brown dwarfs, producing a new class of hypervelocity objects that eventually leaves the host galaxy. {}{Since the clump formation is most vigorous for adiabatic indices $\gamma \gtrsim 5/3$, those adiabatic indices being somewhat unphysical for real stellar progenitors, it may seem as though the production of unbound objects is largely inhibited for realistic TDEs; however, if cooling can significantly decrease the entropy (see below), then the number of clumps could be significantly augmented. Therefore, if there are between $10^{-4}$ and $10^{-5}$ disruptions per galaxy per year, the number of hypervelocity, low-mass objects could significantly outweigh the number of hypervelocity stars.} We plan to perform a more in depth analysis of the detailed properties of the clumps in a future paper.

We also recall that the marginally bound material recedes to very large distances before returning to the black hole. Therefore, similar to the unbound material, the clumps that form in this region of the stream may have time to collapse into much denser objects (e.g., planets). These objects may then be able to survive their plummet back into the tidal region of the black hole (though interactions with the surrounding stellar population may alter their pericenter distances to be larger than the original tidal radius), forming a class of low-mass objects that remain bound to the black hole. Since they would still be very weakly bound, their orbital periods would be anywhere from tens to thousands of years. Furthermore, if the clumps in this region do not become overly dense, they may form weakly-bound clouds that are consistent with those observed near the Galactic Center (e.g., the cloud G2; \citealt{bur12, gil12, gui14a}).

\subsection{Entropy}
In these simulations, the gas maintained approximately constant entropy throughout the entire disruption process. In reality, the gas energy equation will be modified by losses due to radiative cooling and cooling or heating (depending on the optical depth of the stream) due to recombinations, which could significantly alter the equation of state of the gas and affect the nature of the caustic. A more realistic equation of state might therefore be of the form

\begin{equation}
p = S(\mathbf{r},t)\rho^{\gamma} \label{pSgam},
\end{equation}
where $S(\mathbf{r},t)$ is related to the entropy of the gas that is, in general, a function of both space and time. When  $S(\mathbf{r},t)$ is a constant, it is apparent from this expression that, for the same change in density, a smaller adiabatic index results in a correspondingly smaller decrease in the pressure. This scaling then results in a larger cross-sectional radius of the stream, which ultimately enables the debris to better resist the pancake and fragmentation for smaller adiabatic indices. We see, however, that if the entropy decreases with time, then the pressure could decrease faster than would be predicted by an isentropic equation of state. Therefore, if cooling is efficient enough to significantly reduce the entropy of the gas, the pancake could induce fragmentation for $\gamma$ less than 5/3. This result is particularly apparent if we use equation \eqref{pSgam} in equation \eqref{lane-emden}, which shows that the cross-sectional radius scales as

\begin{equation}
H \propto S^{1/2}\rho^{\frac{\gamma-2}{2}},
\end{equation}
and using this relation in equation \eqref{rhostream} yields

\begin{equation}
\rho = \rho_{ad}(r,t)\left(\frac{S}{S_0}\right)^{-n} \label{rhoS}.
\end{equation}
Here $\rho_{ad}$ is the density one obtains for an adiabatic equation of state, given by equation \eqref{rhoeq}, $S_0$ is the original entropy of the gas at the time of stellar disruption, and we recall that $n = 1/(\gamma-1)$. We see that a decrease in the entropy has a more pronounced effect for smaller $\gamma$, meaning that efficient cooling would more easily result in recollapse for softer equations of state. In particular, since $\rho_{ad} \propto t^{-2.4}$ for $\gamma = 1.5$ $(n = 2)$, we would only need $S \propto t^{-0.2}$ to bring the power-law to $\rho \propto t^{-2}$, which would make the stream marginally unstable to gravitational collapse.

\section{Summary and conclusions}
We have shown that a caustic -- a surface where the orbits of the gas parcels comprising the stream of tidally-disrupted debris formally attain infinite density -- results from the impulse approximation applied to $\beta = r_t/r_p \simeq1$ tidal encounters. This pancake is analogous to the one discovered by \citet{car82}; however, in this case the pancake occurs post-periapsis (on the order of an hour after the star reaches pericenter for the disruption of a solar-type star by a $10^6M_{\astrosun}$ hole), and the compression occurs in the plane of the orbit of the stream, which causes the orientation of the pancake to be perpendicular to the plane of the orbit (see Figure \ref{fig:pancakes}). 

In a realistic TDE, the pressure of the gas will prevent the existence of a true caustic. To test the effects of pressure in resisting the pancake, we simulated four tidal encounters between a solar type star ($R_* = R_{\astrosun}$ and $M_* = M_{\astrosun})$ and a $10^6M_{\astrosun}$ hole with the pericenter of the center of mass of the parabolic, stellar orbit at the tidal radius ($\beta = 1$). The simulations differed only in the adiabatic index of the gas, being $\gamma = 1.5$, 5/3, 1.8 and 2, making our parameter space close to that chosen by \citet{lod09}. 

A few hours after disruption, the density of the streams of debris produced by the disruption exhibit anomalous behavior, showing compression-rarefaction oscillations not accounted for by the analytic model (see Figures \ref{fig:rhogamma53} -- \ref{fig:rhomaxvt_allgammas}). We interpret these features as arising from the \ed{combination of the} perpendicular pancake \ed{and self-gravity}, not only because of the temporal coincidence of the two phenomena, but also because the majority of the stream seems to be undergoing a systematic increase in the density at the start of the first compression. This can be seen by noting that the first increase in the maximum stream density starts at a time of roughly an hour after disruption, yet the stream seems to retain its stretched-polytropic structure, predicted analytically, after six hours post-disruption (compare Figures \ref{fig:rhomaxanalytic} and \ref{fig:rhogamma53}). This suggests that a large portion of the stream is being compressed simultaneously and by the same factor, which is predicted for the pancake\ed{; furthermore, this systematic increase in the density was not observed in the test runs that avoided the caustic (see section 4.1)}. By using the periapsis velocities and positions of the gas parcels generated from the {\sc phantom} runs as the initial conditions for an N-body simulation, we also showed that the orbits of the central portions of the stream do tend to form a caustic (Figure \ref{fig:nbodyplot}). This finding is consistent with the fact that the dense, central portions of the star likely retain their structure better until reaching pericenter, thus creating the conditions necessary to form a post-periapsis pancake. On the other hand, the outer, less-dense regions of the envelope are stripped from the star sooner, violating the impulse criterion that they move with the center of mass until reaching pericenter, and thus avoiding the caustic.

The self-gravity of the stream\ed{, supplemented by the caustic,} induces fluctuations on top of the otherwise-smooth, analytically-predicted density profile, as evidenced in Figure \ref{fig:rhoallgammas}. The fact that the analytic predictions match the numerical solutions well means that the stream width is predominantly set by the balance between pressure and self-gravity, and {}{does not undergo any episode of} free expansion {}{immediately after pericenter passage} as expected previously \citep{koc94}. The effects of the caustic are long-lived, and the density profile of the stream evolves for a considerable amount of time after the initial perturbations are imposed. The variations induced by the caustic \ed{and self-gravity} drive deviations from the canonically-assumed $t^{-5/3}$ fallback rate, the power-law being first steeper and then shallower than $5/3$, which can be seen from Figure \ref{fig:mdots}. The peak in the accretion rate is also higher than would be predicted analytically (Figure \ref{fig:mdot_gamma1p8}). 

Remarkably, \ed{the combination of the caustic and self-gravity} can cause the stream to fragment into small-scale, gravitationally-bound clumps if the adiabatic index is high enough. Specifically, for $\gamma = 2$ and $\gamma=1.8$, we find that the stream collapses at a time of a few days and a couple of weeks, respectively. After a relatively short time --- about two weeks for $\gamma=2$ and two months for $\gamma = 1.8$ --- the clump formation stops and the masses of the clumps saturate. For $\gamma = 2$ the average clump mass is $\bar{M}_{c} \simeq 2.6M_{J}$, while that for the $\gamma = 1.8$ run is $\bar{M}_{c} \simeq 0.55M_{J}$, $M_J \simeq 9.54\times10^{-4}M_{\astrosun}$ being the mass of Jupiter. In both of these cases, however, the maximum clump mass is an order of magnitude above the average, showing that there is a large range of clump masses. For $\gamma = 5/3$, the stream does collapse, but the instability is started by small scale noise and so future simulations with realistic perturbations are required (see also \citealt{cou15}). For $\gamma = 1.5$, we find no fragmentation out to a simulated time of ten years post-disruption, suggesting that the stream is gravitationally stable.

The formation of these clumps has a number of interesting repercussions. For one, if an accretion disk has already formed from the tidally-stripped debris, it can intercept one of the infalling clumps and, especially if the clump mass is on the larger side ($\gtrsim 1M_{J}$) of the distribution, significantly augment the accretion rate onto the black hole if the viscous timescale in disk is short (see Figure \ref{fig:mdot_gamma1p8}). Such periodic increases would be seen as variability in the lightcurve of the TDE, consistent with that observed for \emph{Swift} J1644+57 \citep{bur11,lev11,zau11}, and also for the events \emph{Swift} J2058+05 \citep{cen12} and \emph{Swift} J1112.2-82 \citep{bro15}. If an accretion disk has not yet formed, these clumps can be ``redisrupted,'' creating complicated interactions between the incoming and outgoing debris streams. This would then tend to isotropize the accretion process onto the hole and and cause increased variability in the lightcurve of the TDE. The clumps that form in the marginally bound material may have time to condense into more compact objects, such as planets and brown dwarfs, that can survive their eventual return to pericenter, allowing them to remain bound to the hole. The clumps formed in the marginally bound segment of the stream may also form less dense clouds, the likes of which are observed near our own Galactic Center (e.g., G2; \citealt{gil12, bur12}). Finally, the unbound clumps may form a new class of low-mass, hypervelocity objects that make their way out of the host galaxy on timescales of millions of years.

Our results are based on the encounter between a solar-type star and a $10^6M_{\astrosun}$ black hole. In reality, the properties of the star and black hole undergoing a tidal encounter may differ from the fiducial parameters chosen here. However, the existence of the in-plane pancake, and the observational consequences derived therefrom, depends only on the fact that the pericenter distance be comparable to the tidal radius. In particular, if $\beta \gg 1$, the star will be disrupted well before reaching periapsis, while if $\beta \ll 1$ the star will only be partially disrupted, as noted by \citet{gui13}. Interestingly, \citet{gui13} also found that if $0.75 \lesssim \beta \lesssim 0.85$ for a $\gamma = 5/3$ equation of state, the star was initially completely destroyed by a $10^6M_{\astrosun}$ black hole; however, at a time greater than $10^4$ seconds post-disruption, the central portion of the stream recollapsed into a single, massive core, with the outer extremities of the stream remaining as tidal tails. We suggest that the perpendicular pancake pointed out here may \ed{have contributed to this} recollapse, and we plan to further investigate this possibility.

The origin of the pancake can be seen directly from equation \eqref{rdoti}, which shows that the gas parcels comprising the front of the star at the time of disruption are decelerating, while those at the back are accelerating; this results in the eventual merger of the in-plane edges of the stream. The differential acceleration across the star is given by equation \eqref{Deltaphidot}, which shows that the magnitude of the pancake is primarily affected by the properties of the progenitor star. However, the inverse scaling with the black hole mass, although weak, implies that smaller mass black holes lead to a larger differential acceleration and, hence, stronger pancakes.

The pancake alone can \ed{augment the self-gravity to the point where the stream gravitationally fragments} in the cases where $\gamma = 1.8$ and 2, and this result is ultimately related to the fact that larger adiabatic indices result in a decreased resistance to the compression. However, a non-adiabatic equation of state could alter these results quite dramatically. In particular, any cooling would decrease the equilibrium width of the stream, enabling the pancake to leave a much more pronounced effect on the debris. The effects of a time-dependent entropy are also increased for smaller $\gamma$, as is apparent from equation \eqref{rhoS}, meaning that even streams with very low adiabatic indices could collapse if the gas-energy equation were evolved self-consistently. We plan to investigate alternative equations of state in a future paper.

\section*{Acknowledgments}

This work was supported in part by NASA Astrophysics Theory Program grants NNX14AB37G and NNX14AB42G, NSF grant AST-1411879, and NASA's Fermi Guest Investigator Program. CN was supported by NASA through the Einstein Fellowship Program, grant PF2-130098. CN was supported by the Science and Technology Facilities Council (grant number ST/M005917/1). {We used {\sc splash} \citep{pri07} for the visualization.} DJP is supported by a Future Fellowship, FT 130100034, from the Australian Research Council. This work utilized the Complexity HPC cluster at the University of Leicester which is part of the DiRAC2 national facility, jointly funded by STFC and the Large Facilities Capital Fund of BIS. This work used the \texttt{Janus} supercomputer, which is supported by the National Science Foundation (award number CNS-0821794) and the University of Colorado Boulder.  The \texttt{Janus} supercomputer is a joint effort of the University of Colorado Boulder, the University of Colorado Denver, and the National Center for Atmospheric Research. {}{We would also like to thank the anonymous referee for useful comments and suggestions, and in particular for emphasizing the importance of the number of hypervelocity objects that could be generated over cosmological timescales.}

\bsp	
\label{lastpage}
\end{document}